\documentclass[twocolumn]{aastex63}

\usepackage{color}
\usepackage{graphicx}
\usepackage{amsmath, amssymb, bm}
\usepackage{CJK}
\usepackage{xspace}

\definecolor{red}{rgb}{1.0,0.0,0.0}

\newcommand{\uatnum}[1]{\href{http://vocabs.ands.org.au/repository/api/lda/aas/the-unified-astronomy-thesaurus/current/resource.html?uri=http://astrothesaurus.org/uat/#1}{#1}}

\newcommand{\refstar}{\mathrm{star}}
\newcommand{\onstar}{\mathrm{onstar}}
\newcommand{\planet}{\mathrm{planet}}
 
\newcommand{\onplanet}{\mathrm{onplanet}}
\newcommand{\as}{\hbox{$^{\prime\prime}$}\xspace}

\newcommand{\RA}{\mathrm{RA}}
\newcommand{\DEC}{\mathrm{DEC}}

 \newcommand{\cam}{
  Institute of Astronomy, University of Cambridge, Madingley Road, Cambridge CB3 0HA, United Kingdom}
 \newcommand{\lesia}{
  LESIA, Observatoire de Paris, PSL, CNRS, Sorbonne Universit\'e, Universit\'e de Paris, 5 place Jules Janssen, 92195 Meudon, France}
 \newcommand{\ipag}{
  Universit\'e Grenoble Alpes, CNRS, IPAG, 38000 Grenoble, France}
 \newcommand{\mpe}{
  Max Planck Institute for extraterrestrial Physics, Giessenbachstra\ss e~1, 85748 Garching, Germany}
 \newcommand{\mpia}{
  Max Planck Institute for Astronomy, K\"onigstuhl 17, 69117 Heidelberg, Germany}
 \newcommand{\esog}{
  European Southern Observatory, Karl-Schwarzschild-Stra\ss e 2, 85748 Garching, Germany}
 \newcommand{\esoc}{
  European Southern Observatory, Casilla 19001, Santiago 19, Chile}
 \newcommand{\lam}{
  Aix Marseille Univ, CNRS, CNES, LAM, Marseille, France}
 \newcommand{\caltech}{
  Department of Astronomy, California Institute of Technology, Pasadena, CA 91125, USA}
 \newcommand{\stsci}{
  Space Telescope Science Institute, Baltimore, MD 21218, USA}
 \newcommand{\umich}{
  Astronomy Department, University of Michigan, Ann Arbor, MI 48109 USA}
 \newcommand{\monash}{
  School of Physics and Astronomy, Monash University, Clayton, VIC 3800, Melbourne, Australia}
 \newcommand{\cornell}{
  Center for Astrophysics and Planetary Science, Department of Astronomy, Cornell University, Ithaca, NY 14853, USA}
 \newcommand{\cologne}{
  1. Institute of Physics, University of Cologne, Z\"ulpicher Stra\ss e 77, 50937 Cologne, Germany}
 \newcommand{\lisboa}{
  Universidade de Lisboa - Faculdade de Ci\^encias, Campo Grande, 1749-016 Lisboa, Portugal}
 \newcommand{\centra}{
  CENTRA - Centro de Astrof\'{\i}sica e Gravita\c c\~ao, IST, Universidade de Lisboa, 1049-001 Lisboa, Portugal}
 \newcommand{\leiden}{
  Leiden Observatory, Leiden University, P.O. Box 9513, 2300 RA Leiden, The Netherlands}
 \newcommand{\dublin}{
  School of Physics, University College Dublin, Belfield, Dublin 4, Ireland}
 
 \newcommand{\bonn}{
  Max Planck Institute for Radio Astronomy, Auf dem H\"ugel 69, 53121 Bonn, Germany}
 \newcommand{\liege}{
  STAR Institute/Universit\'e de Li\`ege, Belgium}

 \newcommand{\australie}{
  Research School of Astronomy \& Astrophysics, Australian National University, ACT 2611, Australia}
 \newcommand{\exeter}{
  University of Exeter, Physics Building, Stocker Road, Exeter EX4 4QL, United Kingdom}
 \newcommand{\porto}{
  Universidade do Porto, Faculdade de Engenharia, Rua Dr. Roberto Frias, 4200-465 Porto, Portugal}
 \newcommand{\amherst}{
  Five College Astronomy Department, Amherst College, Amherst, MA 01002, USA}
 \newcommand{\ucb}{
  Department of Astronomy, University of California at Berkeley, CA 94720, USA}

\submitjournal{AAS Journals}

\shorttitle{PDS 70 Protoplanets Interferometry}
\shortauthors{Wang et al.}

\begin{document}
\begin{CJK*}{UTF8}{gbsn}

\title{Constraining the Nature of the PDS 70 Protoplanets with VLTI/GRAVITY\footnote{Based on observations collected at the European Southern Observatory under ESO programmes 0101.C-0281(B), 1103.B-0626(A), 2103.C-5018(A), and 1104.C-0651(A).}}

\correspondingauthor{Jason Wang}
\email{jwang4@caltech.edu}

% author list begin
\author[0000-0003-0774-6502]{J. J. Wang (王劲飞)}
\altaffiliation{51 Pegasi b Fellow}
\affiliation{\caltech}
\author[0000-0002-5902-7828]{A.~Vigan}
\affiliation{\lam}
\author[0000-0002-6948-0263]{S.~Lacour}
\affiliation{\lesia}
\affiliation{\esog}
\author{M.~Nowak}
\affiliation{\cam}
\author[0000-0002-5823-3072]{T.~Stolker}
\affiliation{\leiden}
\author[0000-0002-4918-0247]{R.~J.~De Rosa}
\affiliation{\esoc}
\author{S.~Ginzburg}
\altaffiliation{51 Pegasi b Fellow}
\affiliation{\ucb}
\author[0000-0002-8518-9601]{P.~Gao}
\altaffiliation{51 Pegasi b Fellow}
\affiliation{\ucb}
\author{R.~Abuter}
\affiliation{\esog}
\author{A.~Amorim}
\affiliation{\lisboa}
\affiliation{\centra}
% ...
\collaboration{200}{The GRAVITY Collaboration}

\begin{abstract}
We present $K$-band interferometric observations of the PDS 70 protoplanets along with their host star using VLTI/GRAVITY. We obtained $K$-band spectra and 100 $\mu$as precision astrometry of both PDS 70 b and c in two epochs, as well as spatially resolving the hot inner disk around the star. Rejecting unstable orbits, we found a nonzero eccentricity for PDS 70 b of $0.17 \pm 0.06$, a near-circular orbit for PDS 70 c, and an orbital configuration that is consistent with the planets migrating into a 2:1 mean motion resonance. Enforcing dynamical stability, we obtained a 95\% upper limit on the mass of PDS 70 b of 10 $M_\textrm{Jup}$, while the mass of PDS 70 c was unconstrained. The GRAVITY $K$-band spectra rules out pure blackbody models for the photospheres of both planets. Instead, the models with the most support from the data are planetary atmospheres that are dusty, but the nature of the dust is unclear. Any circumplanetary dust around these planets is not well constrained by the planets' 1-5 $\mu$m spectral energy distributions (SEDs) and requires longer wavelength data to probe with SED analysis. However with VLTI/GRAVITY, we made the first observations of a circumplanetary environment with sub-au spatial resolution, placing an upper limit of 0.3~au on the size of a bright disk around PDS 70 b.
\end{abstract}


\section{Introduction} \label{sec:intro}
The process of transforming the dust around stars into mature planetary systems is complex and multifaceted. The initial stages of planet formation are mostly hidden from observations as planets grow from small embryos to large cores to natal protoplanets through processes such as streaming instability, planetesimal accretion, or even gravitational instability \citep{Bodenheimer1974,Pollack1996, Youdin2005}. For gas giants like our own Jupiter, after they have grown large enough to undergo runaway growth, we can begin to indirectly observe them as they carve out gaps and excite density waves in the circumstellar disk as they orbit the star \citep[e.g.,][]{vanderMarel2013, Casassus2013, Perez2014, ALMA2015, Andrews2018}. During this process, the dynamical interactions between the protoplanet and the disk can cause the planet to migrate in the disk \citep{Lin1986, Ward1997,Duffell2014}. In systems with multiple protoplanets, interactions with the disk and planets can excite eccentricities, cause dynamical instabilities, or lock the planets in resonance \citep{Dong2016}. As the protoplanets accrete more material, they grow to detectable levels, and emerge from the shroud of dust and gas that obscured them from direct observations \citep{Zhu2015, GinzburgChiang2019, Szulagyi2019}. After the circumstellar gas disk clears, the gas giant planet formation process is effectively over, leaving behind the many planetary systems we see today.

Catching a glimpse of protoplanets in the process of forming is difficult due to circumstellar and circumplanetary dust shrouding them at the earliest times \citep{Zhu2015, Szulagyi2019}. The distances of nearby systems young enough to still be undergoing planet formation \citep[e.g.][]{Boccaletti2020} are $\gtrapprox 100$~pc making it difficult to spatially resolve them from their circumstellar disks using single-dish 8-10~m telescopes. For both of these reasons, the ability to identify and characterize young protoplanets currently have been limited. Several protoplanets have been reported \citep{Kraus2012, Quanz2013, Biller2014, Reggiani2014, Currie2015, Sallum2015}, but have had their classification questioned \citep{Thalmann2015, Rameau2017, Follette2017, Mendigutia2018, Ligi2018}.

Out of all the sources reported, only the two sources around PDS 70 are undeniably protoplanets in nature. Like many other protoplanet candidates, the star PDS 70 harbors a circumstellar disk with features such as a large gap that may be due to planets in the system \citep{Hashimoto2012, Dong2012, Hashimoto2015}. As part of the SHINE exoplanet survey \citep{Chauvin2017, Vigan2020shine}, PDS 70 b was discovered clearly inside the gap and imaged at multiple wavelengths unlike other protoplanet candidates, making it easy to rule out confusion with circumstellar disk features \citep{Keppler2018, Muller2018}. PDS 70 c could not be confidently identified as a protoplanet initially due to the fact it appeared adjacent to the rim of the circumstellar disk in projection. It was discovered with H$\alpha$ imaging \citep{Haffert2019}, which took advantage of the fact that only protoplanets and their host stars that are actively accreting material are hot enough to emit strong atomic hydrogen emission lines. The protoplanet nature of both planets was confirmed by their strong H$\alpha$ detections that imply mass accretion rates of at least $10^{-8}~M_\textrm{Jup}/\textrm{yr}$ \citep{Wagner2018, Haffert2019}. Assuming their orbits are coplanar with the circumstellar disk that has been well characterized in the near-infrared and mm \citep{Keppler2018, Keppler2019, Francis2020}, the planets are near the 2:1 period commensurability \citep{Haffert2019, Wang2020}. Dynamical studies have shown that having these two planets in mean-motion resonance (MMR) would be stable and could create the disk features we see \citep{Bae2019,Toci2020}. However, with a short orbital arc and uncertainties of several mas, their exact orbit remains uncertain \citep{Wang2020}.

Owing to their protoplanetary nature, it is currently inconclusive what emission we are seeing from the planets and their circumplanetary environments. Current low-resolution spectroscopy and photometry from 1-5 $\mu$m point to emission that is very dusty with only one tentative water absorption feature for PDS 70 b \citep{Muller2018, Haffert2019, Mesa2019, Christiaens2019MNRAS, Wang2020, Stolker2020}. The cause of the dusty spectrum could be due to high level hazes in the atmosphere, the accretion of material onto the planet, or a circumplanetary or circumstellar disk obscuring the planet, with recent analysis favoring accreting material being responsible \citep{Wang2020,Stolker2020}. The spectral characterization of PDS 70 c has been especially challenging, requiring high angular resolution imaging and disk modeling to properly extract photometry from the protoplanet \citep{Haffert2019, Mesa2019, Wang2020, Stolker2020}. Despite having a relatively large wavelength coverage, the emission from both planets was found to still be consistent with a single blackbody, with no support from the data for using more sophisticated models \citep{Wang2020, Stolker2020}. However, it is unlikely that the true emission from these protoplanets are blackbodies. As these planets are accreting, there also should be dust in their circumplanetary environments. There has been tentative evidence for a circumplanetary disk (CPD) around PDS 70 b with $K-$ and $M$-band excess \citep{Christiaens2019APJ, Stolker2020}, and a significant ALMA detection of dust at the location of PDS 70 c \citep{Isella2019}.

Long-baseline optical interferometry allows us to combine multiple single-dish telescopes together to achieve an order of magnitude boost in angular resolution, important for discerning protoplanets from circumstellar and circumplanetary dust \citep{Wallace2019}. Recently, the GRAVITY interferometer at VLTI made the first direct detection of an exoplanet with optical interferometry using its pioneering phase-referenced dual-field mode \citep{GRAVITY2019}. This mode has shown GRAVITY can achieve astrometric precisions down to 50~$\mu$as and obtain high signal-to-noise $K$-band spectra of exoplanets at R$\sim$500 \citep{2020A&A...633A.110G, Molliere2020, Nowak2020b}.

In this work, we will leverage the superior angular resolution of GRAVITY combined with its ability to distinguish coherent and incoherent emission to study the PDS 70 protoplanets. In Section \ref{sec:obs}, we describe the observations made of PDS 70 b and c as well as its host star, the data reduction, and spectral calibration. We then fit the orbit of both planets and make dynamical mass constraints based on stability arguments in Section \ref{sec:orbit}. In Section \ref{sec:sed}, we fit multiple atmospheric models, explore the evidence for extinction and circumplanetary disk emission, discuss the nature of the photospheric emission from these protoplanets, and place limits on Br$\gamma$ accretion signatures. We also use the long-baseline data from GRAVITY to attempt to resolve the circumplanetary environment in Section \ref{sec:circum}. Finally we offer some concluding thoughts in Section \ref{sec:conclusions}.

\section{Observations and Data Reduction} \label{sec:obs}

\subsection{GRAVITY Observations}

\begin{table*}
  \begin{center}
    \caption{Log of the observations.}
    \begin{tabular}{c|c|c|c|c|c|c|c|c}      \hline \hline
      Target & Date & UT Time &  Resolution & \multicolumn{2}{c|}{Nexp/NDIT/DIT\tablenotemark{a} }  & airmass & $\tau_0$ & seeing \\
      \hline
       &  & & &  Planet & Star &  &  & \\
      \hline \hline
    PDS\,70\ A & 2018-06-25 & 01:13:21-01:30:12 & MEDIUM & --- & 3/30/10\,s & 1.05-1.07 &  2.0-2.7\,ms & 0.55-0.80\as \\
    HD\,124058 & 2018-06-25 & 01:44:45-02:01:42 & MEDIUM & --- &  3/30/10\,s  &1.07-1.09 & 2.0-2.7\,ms &0.55-0.80\as  \\
      \hline
PDS\,70\ b & 2019-07-16 & 00:09:49-01:01:39 & MEDIUM & 3/12/60\,s & 3/64/1\,s & 1.07-1.15 & 1.7-2.3\,ms & 0.87-1.32\as \\
PDS\,70\ b & 2020-02-10 & 07:09:28-08:38:48 & MEDIUM & 3/8/100\,s & 6/8/10\,s & 1.05-1.21 & 11.0-21.4\,ms & 0.61-0.73\as \\
\hline
PDS\,70\ c & 2019-07-19 & 00:11:46-01:19:17 & LOW & 7/16/30\,s & 8/64/1\,s & 1.08-1.21 & 3.7-5.5\,ms & 0.72-1.2\as \\
PDS\,70\ c & 2020-02-10 & 07:29:01-09:13:45 & MEDIUM & 4/8/100\,s & 6/8/10\,s & 1.05-1.21 & 11.0-21.4\,ms & 0.61-0.73\as \\
      \hline \hline
    \end{tabular}
    \label{tab:log}
    \tablenotetext{a}{Nexp is the number of exposures; NDIT is the number of sub-integrations; DIT is the detector integration time. The three values can be multiplied together for the total integration time. }
  \end{center}
\end{table*}

\begin{table}
  \begin{center}
    \begin{tabular}{c|c|c|c|c|c|c}
      \hline
      \hline
       & MJD & $\Delta\RA$ & $\Delta\DEC$ & $\sigma_{\Delta\RA}$ & $\sigma_{\Delta\DEC}$ & $\rho$ \\
      & (days) & (mas) & (mas) & (mas) & (mas) & - \\
      \hline
b & 58680.032 & 102.61 & -139.93 & 0.09 & 0.24 & 0.39 \\
b & 58889.341 & 104.70 & -135.04 & 0.09 & 0.11 & -0.89 \\
      \hline
c & 58683.034 & -214.95 & 32.22 & 0.13 & 0.13 & 0.28 \\
c & 58889.353 & -214.30 & 27.19 & 0.07 & 0.16 & -0.72 \\
      \hline
      \hline
    \end{tabular}
    \caption{Relative astrometry of PDS 70\,b and c. Due to the interferometric nature of the observations, a correlation coefficient $\rho$ is required to properly describe the confidence intervals, which are not aligned on the sky coordinates. The covariance matrix can be reconstructed using ${\sigma_{\Delta\RA}}^2$ and ${\sigma_{\Delta\DEC}}^2$ on the diagonal, and $\rho\sigma_{\Delta\RA}\sigma_{\Delta\DEC}$ off-diagonal.}
    \label{tab:astrometry}
  \end{center}
\end{table}

The observations were carried out using the GRAVITY instrument \citep{2017A&A...602A..94G} on the VLTI using the four Unit Telescopes (UT). The log of the observations are given in Table~\ref{tab:log}. Atmospheric conditions ranged from very good (atmospheric coherence time $\tau_0 = 20$\,ms) to average ($\tau_0 \approx 2$\,ms). The first observation, in 2018, is a classical interferometric observation of the star (PI M.~Benisty, ID 0101.C-0281). The 2019 observations were obtained as a backup of the AGN large program (PI E.~Sturm, ID 1103.B-0626, for PDS\,70\,b), and from director discretionary time (PI A.~Vigan, ID 2103.C-5018, for PDS\,70\,c). Last, the 2020 observations were obtained as part of the ExoGRAVITY large program (PI S.~Lacour, ID 1104.C-0651).

The 2018 observations were carried out using the single-field on-axis mode. On-axis means the beam splitter was used \citep{2014SPIE.9146E..23P}, therefore sending 50\% of the flux to the fringe tracker \citep{2019A&A...624A..99L} and 50\% on the science channel. Single-field means the fringe tracker and the science fibers observed the same object: in this case, the star PDS\,70\ A. The observations were followed by observations of the calibrator HD\,124058. The reduction of this dataset was standard using the ESO GRAVITY pipeline\footnote{url: \url{https://www.eso.org/sci/software/pipelines/gravity}} \citep{2014SPIE.9146E..2DL}.

The observations of the exoplanets used the dual-field on-axis mode. Dual-field means the fringe tracker and the science fibers observed different objects. Thanks to the splitter, the fringe tracker observed the star for phase referencing, and the science fiber observe the planet. Non-common path phase aberrations were calibrated by interleaving the observation of the planet with single-field on-axis observations.

\subsection{Reduction of Relative Astrometry}
\label{sec:astrometry}

The coherent flux was extracted following a standard procedure with the ESO GRAVITY pipeline. From this first step, we obtain $V_\onplanet(b,t,\lambda)$ and $V_\onstar(b,t,\lambda)$, the coherent flux observed on the star and the planet as a function of baseline $b$, time $t$, and wavelength $\lambda$.

The removal of stellar contamination was performed during a second step. The details of the computation is described in detail in Appendix A of \citet{2020A&A...633A.110G}.  The code is available as a Python library developed by our team\footnote{Software available on GitHub upon request.}. The main objective of the algorithm is to calculate $R(\lambda,b,t)$, the ratio of the uncontaminated coherent flux between the star and planet:
\begin{equation}
R(b,t,\lambda,)= \frac{ \gamma_{\planet} }{\gamma_{\refstar}} \ \frac{V_\planet(b,t,\lambda)}{V_\refstar(b,t,\lambda)}
\end{equation}
where $V_\refstar(b,t,\lambda)$ and $V_\planet(b,t,\lambda)$ are the coherent flux of both objects in the absence of stellar speckle. The additional term $\gamma$ comes from the fact that the science fiber is not exactly positioned at the location of the target in the focal plane (see Appendix~\ref{sec:losses}). The astrometry was obtained from the argument of $R$, the ratio of the coherent flux:
\begin{equation}
\Phi_R(b,t,\lambda)=- \frac{2 \pi}{\lambda} (\Delta \RA\, u + \Delta \DEC\, v)
\end{equation}
in which $(u,v)$ are the coordinates in the frequency domain and $(\Delta \RA,\Delta \DEC)$ are the sky-coordinates of the planet relative to the star.

The values, obtained by $\chi^2$ minimization, are given in Table~\ref{tab:astrometry}. The error bars given here correspond to the precision of the measurement. They are estimated from the scatter of the astrometric values (between each file, or by splitting the files into independent measurements). The typical precision is  $~100\,\mu$as. The systematic errors, which are ultimately limiting the accuracy of the astrometry, are theoretically smaller. They were estimated to be 16.5\,$\mu$  \citep{2014A&A...567A..75L}.

\subsection{Reduction of Spectra and Calibration}
\label{sec:redSpectra}

\begin{figure}
    \centering
    \includegraphics[width=0.49\textwidth]{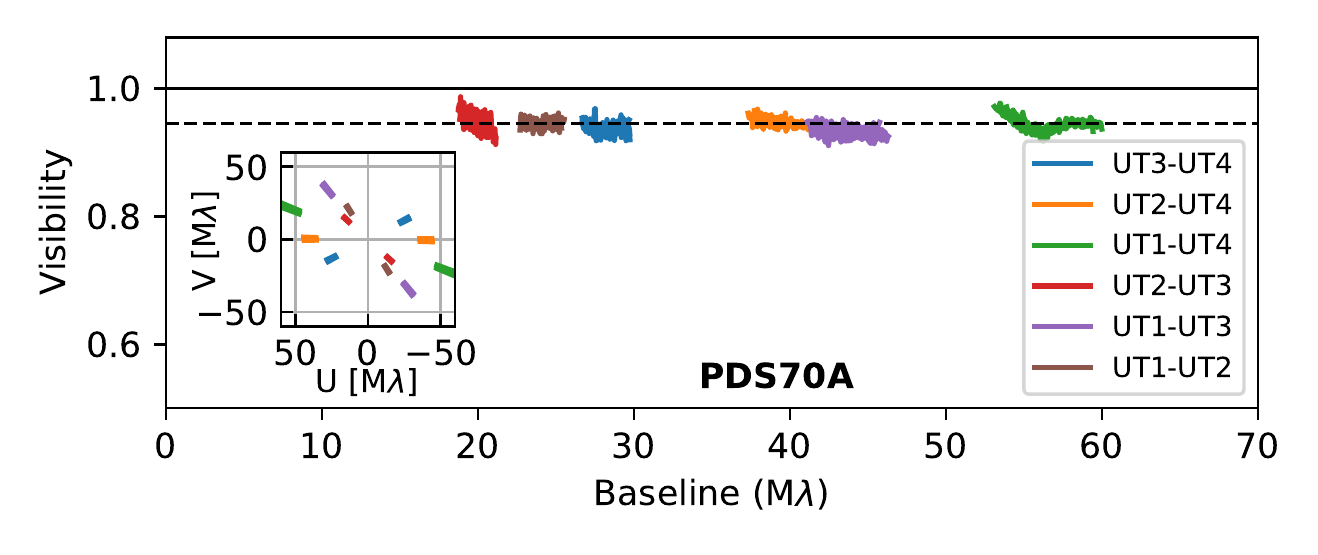}
    \caption{Plot of the visibilities of PDS\,70\ A. The dashed line is the average of the visibility, equal to $94\pm1\%$. The inset panel represents the u-v coverage of the observations.}
    \label{fig:PDS70A}
\end{figure}

The coherent fluxes $V_\refstar(b,t,\lambda)$ and $V_\planet(b,t,\lambda)$  are not normalized and therefore include the shape of the spectrum. To extract the spectrum $F_\planet(\lambda)$, we assumed the planet to be unresolved. 
The amplitude of coherent flux of the planet is then equal to the planetary flux:
\begin{equation}
  V_\planet(b, t, \lambda) = F_\planet(\lambda) \exp(i \Phi(b,t,\lambda)) \,,
  \label{eq:visPlanet}
\end{equation}
with the phase derived from the astrometry obtained as described in the previous section.

The star itself has an angular size much smaller than 0.1~mas and therefore was not resolved by our observations. However, the visibilities are still below one because the interferometer did partially resolve the hot inner disk:
\begin{equation}
  V_\refstar(b, t, \lambda) = F_\refstar(\lambda)J_\refstar(b, t, \lambda)
\end{equation}
where  $J_\refstar(b, t, \lambda)$ is the visibility drop due to resolving the central source (star and inner disk). Hence:
\begin{equation}
  F_\planet(\lambda) =   \frac{ \gamma_{\refstar} F_\refstar(\lambda) }{ \gamma_{\planet}}
    < \frac{R(\lambda,b,t) J_\refstar(b, t, \lambda) }{ \exp(i \Phi(b,t,\lambda))  } >_{b,t} \,.
  \label{eq:Rspectra}
\end{equation}
The notation $<>$ corresponds to the mean notation: the coherent flux ratio is averaged over $b$ and $t$. The injection efficiencies $\gamma_\refstar$ and $\gamma_{\planet}$ are real numbers between 0 and 1, where 1 is the theoretical maximum. The value of $\gamma_{\planet}$ was nearly 1 for most epochs except for the first epoch of PDS\,70\ c when we did not know the precise position of the planet. During this observation, the fiber was pointing 16.5\,mas away from the planet, which gave an injection efficiency of $\rho=0.84$. The calculation of this term is given in Appendix \ref{sec:losses}.

To obtain $J_\refstar(b, t, \lambda)$, we used the observation of PDS\,70\ A, calibrated by the star HD\,124058 (assuming a diameter for the calibrator of $0.136\pm0.002$\,mas).
 The obtained visibilities are showed in  Figure~\ref{fig:PDS70A}. They are mostly consistent with a single constant:
 \begin{equation}
  J_\refstar(b, t, \lambda) = \textrm{cst}\,.
 \end{equation}
The value of the constant does depend on the normalization of the stellar flux. We found that $94 \pm 1$\% of the flux on-axis of the star is unresolved, independent of baseline. That is, 6\% of the flux came in excess emission from the hot inner circumstellar disk that is resolved with GRAVITY. An inner circumstellar disk has been predicted from SED analysis \citep{Hashimoto2012, Dong2012, Long2018}, detected in scattered light \citep{Keppler2018, Mesa2019}, and resolved in the mm \citep{Francis2020}.

Because the inner disk is resolved, we cannot simply use the 2MASS $K$-band photometry of the system \citep{Cutri2003} to normalize the stellar model of the star to calibrate the planetary spectra, since it would include excess emission from the circumstellar disk, which is not part of $F_\refstar(\lambda)$. Fortunately, the combined scattered light and thermal emission from the inner disk at shorter wavelengths contributes $\lesssim$5\% of the total flux from the system \citep{Dong2012}, comparable to the 1$\sigma$ errors on the stellar photometry.
Therefore, we used a BT-NextGen stellar atmosphere \citep{Allard2012} determined from a joint evolutionary-atmospheric model fit to literature optical and near-infrared photometry using the procedure described in \citet{Wang2020}. We did not impose a prior on the effective temperature as in \citet{Wang2020}, but all other aspects of the fit were the same. From the resulting posterior distributions of the fitted and derived parameters, we measured an age of $8\pm1$\,Myr, a mass of $0.88\pm0.02$\,M$_{\odot}$, an effective temperature of $4109^{+36}_{-30}$\,K (higher than the spectroscopically derived value of $3972\pm36$\,K; \citealp{Pecaut:2016fu}), and a surface gravity of $\log g = 4.23\pm0.02$\,[dex]. The mass estimated from SED fiting is significantly higher than previous literature estimates from SED fitting ($0.76\pm0.02$\,M$_{\odot}$; \citealp{Muller2018}), but in better agreement with the dynamical mass estimates from circumstellar disk modeling ($0.875\pm0.03$\,M$_{\odot}$; \citealp{Keppler2019}) and from the orbit fits presented in Section \ref{sec:orbit}. A synthetic stellar spectrum was computed by randomly drawing 200 samples from the Markov Chain Monte Carlo (MCMC) chain, taking the median and standard deviation of the flux at each wavelength as the adopted spectrum and corresponding uncertainties, respectively. We used this spectrum as the spectrum of the star ($F_\refstar$) to calibrate our planet spectra. The statistical uncertainties in the model stellar spectrum are much lower than the uncertainties on the planetary spectra, but do not include any estimate of systematic errors in the stellar models. 

Comparing our stellar model to the 2MASS $K$-band photometry, we also measured a significant $K$-band excess of $14\pm3$\,\%, caused by emission from circumstellar material close to the star. This is much greater than the 6\% excess emission resolved by GRAVITY. As the stellar variability is dominated by the rotation modulation of the star \citep{Thanathibodee2019}, it is unlikely to be responsible for this disagreement, as the photometric measurements used in the stellar SED fit should average over this variability. Rather, the GRAVITY observations probe spatial scales of 0.2-0.5~au, which are right in the middle of the spatial extent of the inner disk based on models from \citet{Dong2012} and \citet{Long2018}. The remaining $\sim$8\% could be emitted closer in to the star where it is not resolved by GRAVITY, or further out at larger spatial scales that are outside the field of view of GRAVITY ($\sim$50 mas). We did not see significant change in the visibilities over the baselines observed, indicating the emission must be significantly closer in or significantly further out. Inner disk models have the inner edge of the disk at 0.05~au so there could be emission that is a factor of $\sim$3 closer in \citep{Dong2012, Long2018}. On the other hand, \citet{Keppler2018} detected the inner disk in polarized light with VLT/SPHERE and found the inner disk could be extended out to 20 au, and \citet{Francis2020} found an outer cutoff of 10~au at mm wavelengths with ALMA. This also agrees with disk-modeling analysis that found the inner disk to end at 15 au \citep{Long2018}. Any emission outside of 6~au would not have been seen by GRAVITY (would not couple into the single mode fibers). The location of the emission affects our photometric calibration as emission at larger separations do not need to be included in $F_\refstar$ whereas emission unresolved by GRAVITY needs to be included in $F_\refstar$ which ultimately changes the absolute brightness of the planets. Without further information at the moment, we assumed that half of the remaining 8\% excess emission comes from the star. For simplicity, we assumed that it has the same spectrum as the star, so we essentially multiplied our model stellar spectrum by 1.04. A 4\% change is already much smaller than the 1$\sigma$ uncertainties on the planet spectra, so the exact scale factor and spectral shape of the excess dust emission should negligibly affect our results.

Since the star is young and accreting with clear H$\alpha$ emission \citep{Thanathibodee2019}, line emission from the star could affect the calibration of the planetary spectra. However, the stellar Pa$\beta$ line has not been detected \citep{Long2018}. Given that the Br$\gamma$ line is expected to be even weaker, and given that it should be unresolved, the impact of the Br$\gamma$ line on a single spectral element of our planetary spectra should be negligible given the relatively large error bars of a single spectral channel ($\sim$15\% of the total flux). Similarly, \citep{Long2018} did not find appreciable CO line emission in the $K$-band, which too should be unresolved in our data, and thus have a negligible impact on our spectrum. Thus, we concluded that our omission of emission lines in our model stellar spectrum is a reasonable approximation.

\begin{figure*}
    \centering
    \includegraphics[width=\textwidth]{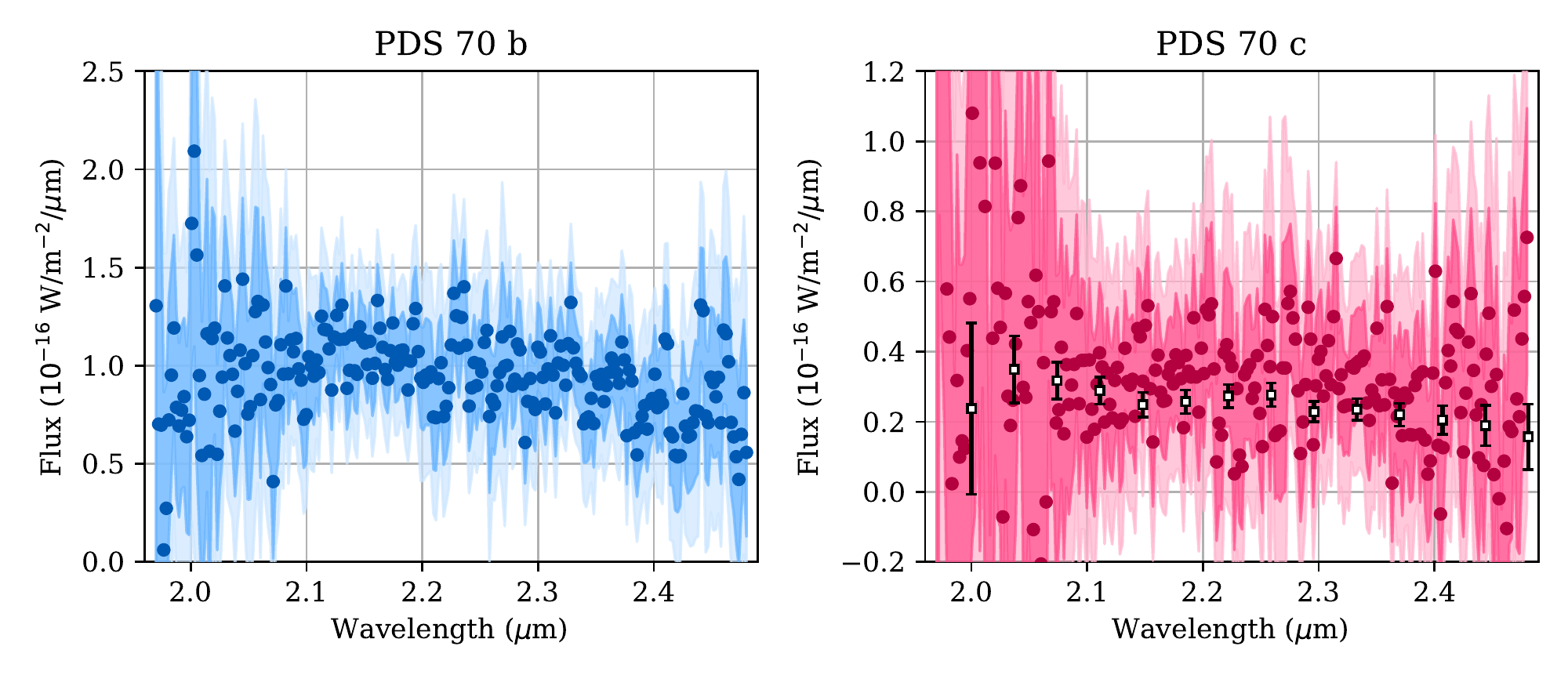}
    \caption{GRAVITY $K$ band spectra of PDS 70 b (left in blue) and PDS 70 c (right in red). In both panels, The circle points denote the estimated flux in each spectral channel in MEDIUM resolution mode. The darker and lighter shaded regions denote the 1$\sigma$ and 2$\sigma$ confidence intervals of the MEDIUM resolution data, accounting for the estimated correlation between neighboring spectral channels. For PDS 70 b, both epochs of data were combined together to create a single spectrum. For PDS 70 c, the LOW resolution epoch is plotted separately as the white squares with black error bars. }
    \label{fig:spectrum}
\end{figure*}

With the coherent flux ratio $R(\lambda,b,t)$ and this model spectrum of the star $F_\refstar(\lambda) $ scaled by 1.04, Eq.~(\ref{eq:Rspectra}) gave us the
 spectra for both planets. The resulting spectrum as well as the uncertainties are plotted in Figure \ref{fig:spectrum}.

\subsection{Reanalysis of SPHERE IFS PDS 70 c data}

\begin{figure}
    \centering
    \includegraphics[width=0.49\textwidth]{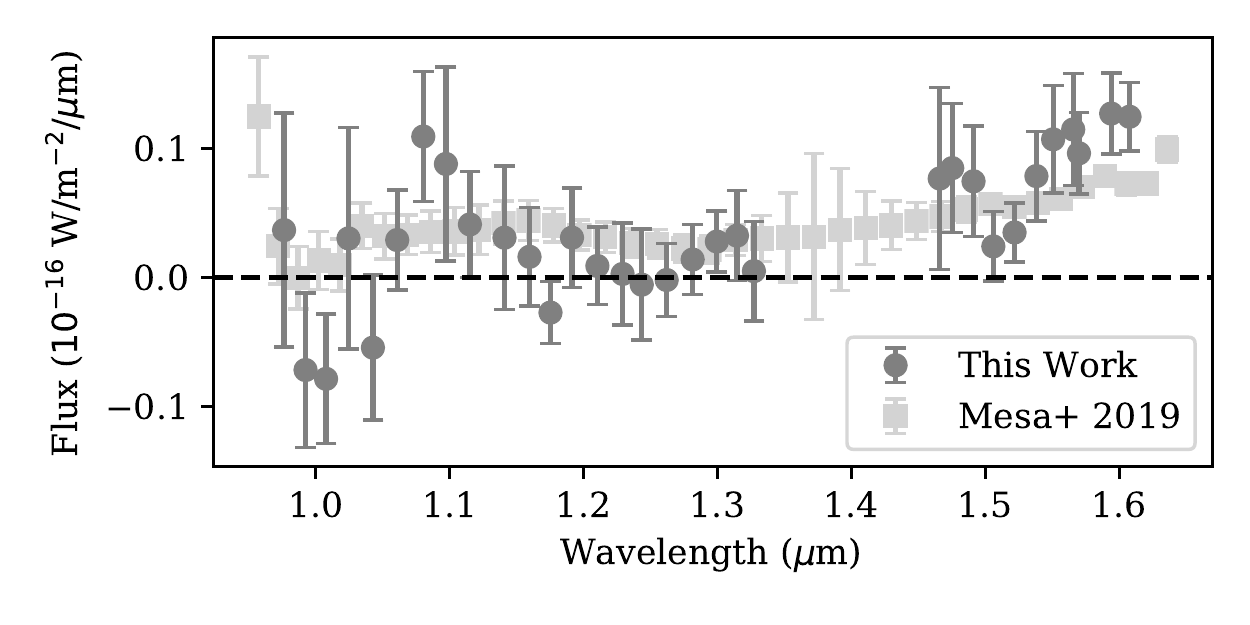}
    \caption{Reanalysis of the PDS 70 c SPHERE IFS spectrum. The dark gray circles are the spectral channels extracted in this work, whereas the light gray squares are from \citet{Mesa2019}. We found larger error bars per spectral channel and did not detect the planet shortwards of $H$-band.}
    \label{fig:c_ifs_spec}
\end{figure}

We also reanalyzed the SPHERE IFS data on PDS 70 c published in \citet{Mesa2019}. Specifically, we reanalyzed the data from 2018-02-24, which were obtained in exquisite conditions (0.40\as seeing). The data were acquired with SPHERE \citep{Beuzit2019} in its \texttt{IRDIFS-EXT} mode where IFS \citep{Claudi2008} and IRDIS \citep{Dohlen2008} observe in parallel, with IFS covering the $YJH$ bands and IRDIS in $K1$ and $K2$ band \citep{Vigan2010}. The data were collected with the apodized pupil Lyot coronagraph \citep{Carbillet2011,Guerri2011} in its \texttt{N\_ALC\_YJH\_S} configuration optimized for the $H$ band. The raw data were preprocessed using the \texttt{vlt-sphere}\footnote{\url{https://github.com/avigan/SPHERE}} open-source pipeline \citep{Vigan2020ascl} to produce calibrated $(x,y,\lambda)$ data cubes of coronagraphic images and off-axis reference PSFs.

Stellar PSF subtraction and spectral extraction of PDS 70 c was done using \texttt{pyKLIP} version 2.1 \citep{Wang2015}. We used angular differential imaging \citep[ADI;][]{Liu2004,Marois2006} and spectral differential imaging \citep[SDI;][]{Sparks2002} to build up a model of the stellar PSF. We used any frame where PDS 70 c moved by one pixel due to ADI and SDI to calculate the principal components to model the star. We used the first 10 principal components, as this gave us the best signal-to-noise on the planet. We used the forward modeling framework described in \citet{Pueyo2016} and \citet{Greenbaum2018} to measure the spectrum of PDS 70 c. We injected eight simulated planets at the same separation but at different azimuthal positions as PDS 70 c, measured their spectra in the same way, and used the scatter in their measured spectra to estimate the uncertainties on the spectrum of PDS 70 c.

Due to the fact the planet is adjacent to the edge of the circumstellar disk, there is concern that the spectral extraction is biased by the disk even with forward modeling. To assess this, we injected five simulated planets at similar separations as PDS 70 c but at other azimuthal positions in the image where the simulated planets would be adjacent to the disk edge and computed the average bias in the flux after spectral extraction. We found and corrected for biases that were at most the size of the 1$\sigma$ uncertainty of each spectral channel. We also verified that the scatter in flux between these five planets is consistent at the 20\% level with the uncertainty we estimated for PDS 70 c in the previous paragraph. In both cases, the errors in the spectral measurements are dominated by the low signal-to-noise of the planet, and not by systematics due to the presence of disk signal. 

We found that PDS 70 c is only detected in the last 12 spectral channels of the IFS data (i.e., $H$-band). Unlike \citet{Mesa2019}, the $Y-$ and $J-$ band spectra are consistent with non-detections, and the scatter we measured at those wavelengths is due to noise. We note that the PDS 70 c spectrum from \citet{Mesa2019} only significantly deviates from their median spectrum of the circumstellar disk in the $H$-band (see their figure 5, panel c), which may indicate their PDS 70 c spectrum at $Y$ and $J$ bands are contaminated by the disk. Our measured spectrum appears to be less affected by the disk likely because we used a more aggressive stellar PSF-subtraction routine that filtered out more disk signal. We also found larger uncertainties per spectral channel on average. Our extracted spectrum of PDS 70 c is plotted in Figure \ref{fig:c_ifs_spec}. Both reductions have indications of correlated noise, as neighboring spectral channels have less scatter than the error bars would imply for uncorrelated spectral channels.

\section{Orbital Dynamics} \label{sec:orbit}

\begin{figure*}
    \centering
    \includegraphics[width=\textwidth]{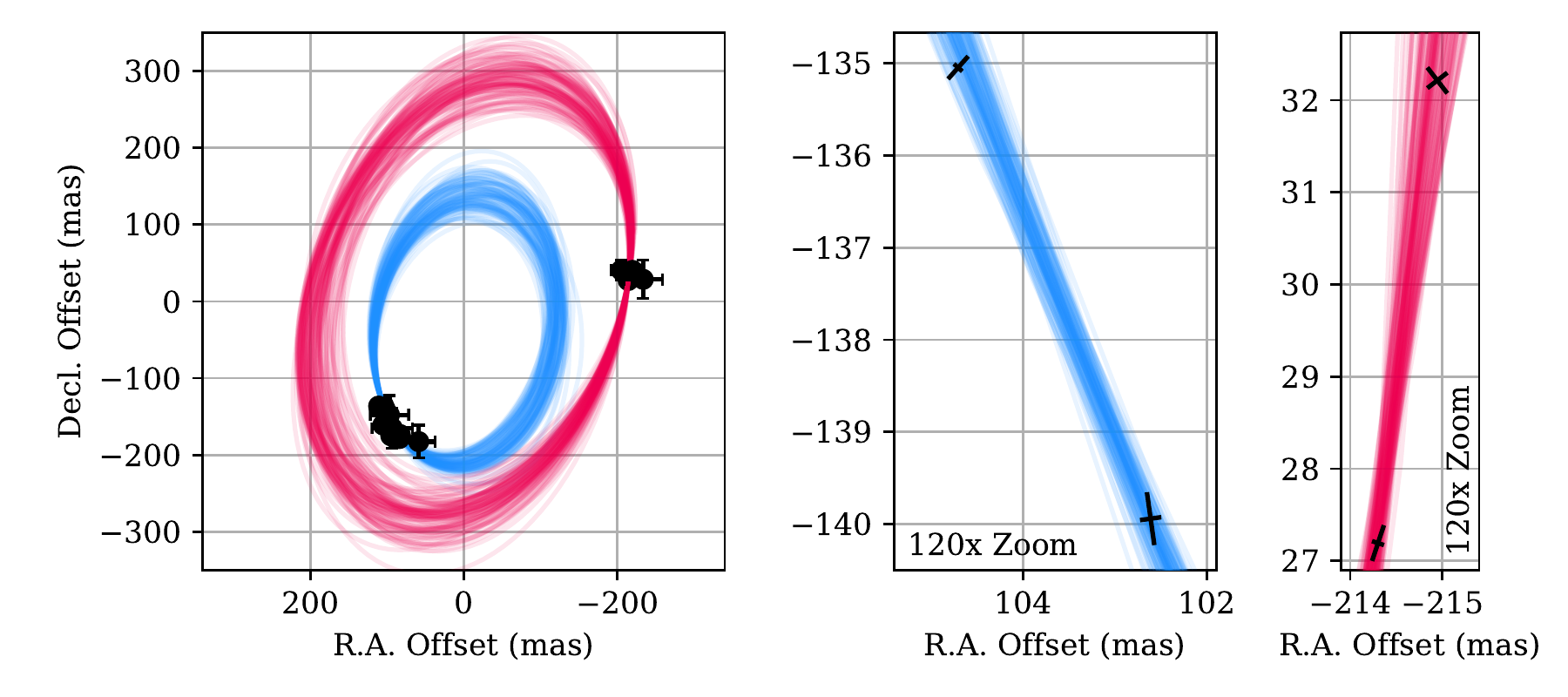}
    \caption{Orbits of PDS 70 b \& c projected onto the sky. On the left, 100 orbits randomly drawn from the posteriors are plotted in blue and red for PDS 70 b and c respectively, and measured astrometry from imaging (doesn't include GRAVITY measurements) are plotted in black. On the center and right are zoomed-in versions of the same plot showing the GRAVITY astrometry for PDS 70 b and c respectively. The tilted error bars in the GRAVITY data represent the principal axes of the error ellipse. }
    \label{fig:skyorbit}
\end{figure*}

\subsection{Orbit Fitting}
With only two epochs of GRAVITY measurements for each planet, we were able to constrain the positions and velocities of PDS 70 b and c with 100$\,\mu$as precision (see Table \ref{tab:astrometry}, but the accelerations and ultimately the orbital elements of the planet are still limited by the precision of astrometry from single dish telescopes. We supplemented the GRAVITY astrometry with imaging astrometry from \citet{Muller2018}, \citet{Mesa2019}, \citet{Haffert2019}, and \citet{Wang2020} (see Table \ref{table:lit_astrom} in Appendix \ref{sec:lit_data}). 

To define the orbit, we used the following orbital parameters: semi-major axis ($a$), eccentricity ($e$), inclination ($i$), argument of periastron ($\omega$), longitude of the ascending node ($\Omega$), epoch of periastron in units of fractional orbital period ($\tau$), system parallax, and the component masses of each body \citep{Blunt2020}. We defined the orbital elements for each planet in Jacobi coordinates as they vary less when accounting for the effects of multiple planets (see next paragraph). The reference epoch for $\tau$ for both planets was MJD 55,000 (2009-06-18). To keep the orbits realistic, we used the same priors as \citet{Wang2020} to impose PDS 70 b and c have non-crossing orbits as well as near coplanarity of the planets and circumstellar disk. We rejected all orbits for which periastron of PDS 70 c is inside of apastron of PDS 70 b. We also applied a Gaussian prior centered at 0 with a standard deviation of 10 degrees on the coplanarity of planet b with the disk, planet c with the disk, and planet b with planet c. We fixed the disk plane to $i = 128.3^\circ$ (equivalent to $i = 51.7^\circ$ but for clockwise orbits) and $\textrm{PA} = 156.7^\circ$ based on \citet{Keppler2019} measurements of the outer disk. The only prior we changed is the one on the stellar mass: we used a Gaussian prior centered at 0.88 $M_\odot$ with a standard deviation of 0.09 $M_\odot$ based on our stellar SED fit in Section \ref{sec:redSpectra} but with 10\% errors to account for model systematics. All our priors are listed in Table \ref{table:orbit}.

We are sensitive to perturbations on the visual orbit of one of the planets around the star due to the other planet. Essentially, our visual orbits are defined by the relative separation between the planet and the star. A second planet, to first order, perturbs the star's position and causes the measured distance between the first planet and the star to change, creating epicycles in the visual orbit  (note that this effect is different from direct planet-planet gravitational interactions, which we will not consider and are much smaller in amplitude). We followed the prescription defined in \citet{BrandtG2020} where only the perturbations of inner planets are accounted for. Thus, the visual orbit of PDS 70 c relative to the star is sensitive to the orbit and mass of PDS 70 b, but not the other way around.  For a Jupiter-mass planet at 20~au, the peak-to-valley amplitude of this perturbation is 400 $\mu$as. To properly model the GRAVITY astrometry, we needed to account for this effect. However, we note that with only two GRAVITY epochs per planet, we did not constrain the masses. Simply, the orbital elements we inferred would have been different if we assumed the planets were massless rather than Jovian mass. In this work, we added uniform priors on planet mass between 1 and 15 Jupiter masses for each planet. Even though recent work \citep[e.g.,][]{Wang2020, Stolker2020} inferred masses closer to 1 $M_\textrm{Jup}$ than 15 $M_\textrm{Jup}$, we purposely extended the prior range to higher masses to assess if we could rule out high-mass solutions via dynamical stability arguments.

The orbital parameters were inferred by Bayesian parameter estimation using an unreleased version of the \texttt{orbitize!} package with commit id \texttt{83356d9} \citep{Blunt2020}, which uses the parallel-tempered affine-invariant sampler \texttt{ptemcee} \citep{ForemanMackey2013,Vousden2016}. This version of \texttt{orbitize!} automatically handles the covariances of the uncertainties in R.A. and decl. that result due to the u-v coverage of the observations. We ran the sampler using 20 temperatures, 1000 walkers per temperature, and 100,000 steps per walker. Convergence was assessed by visual inspection of the walker chains, and by checking that we ran the sampler for at least 100 autocorrelation times. We also accounted for the perturbations on the visual orbits of each planet due to the other planet in the system as described in the previous paragraph. The posterior was formed using the last 40,000 steps from each walker at the lowest temperature. The visual orbit for PDS 70 b is plotted in Figure \ref{fig:skyorbit} and the posterior credible intervals (CIs) are listed in Table \ref{table:orbit}.

With the new GRAVITY data, we constrained the semi-major axis of PDS 70 b ($a_b$) to $\pm2$ au (95\% credible interval), but $a_c$ to only $\pm5$ au for the same credible interval. Perfectly circular orbits for PDS 70 b are disfavored by the current data whereas PDS 70 c is consistent with circular orbits. The period ratio of PDS 70 c to PDS 70 b is $2.13^{+0.27}_{-0.24}$ (68\% credible interval; $2.13^{+0.56}_{-0.45}$ for the 95\% credible interval), putting it near the 2:1 mean-motion resonance (MMR) as has been proposed by \citet{Haffert2019} and \citet{Bae2019}. For the first time, we are able to strongly disfavor all other first-order mean-motion resonances such as the 3:2 and 4:3 MMR. Given these planets are thought to be Jovian mass \citep{Wang2020,Stolker2020}, if they are locked in MMR, it would likely require the strength of a first-order resonance \citep[e.g.,][]{Andre2016}. Thus, the 2:1 MMR is the single likely candidate for orbital resonance.

\begin{deluxetable*}{c|c|c|c}
\tablecaption{Orbital Parameters for PDS 70 b and c  \label{table:orbit}}
\tablehead{
Orbital Element  & Prior & Near-Coplanar & Dynamically Stable
}
\startdata
 $a_b$ (au)               & LogUniform(1, 100)\tablenotemark{a} & $20.1^{+0.9~(+1.9)}_{-1.0~(-1.9)}$ & $20.8^{+0.6~(+1.3)}_{-0.7~(-1.1)}$ \\
 $e_b$                    & Uniform(0, 1)\tablenotemark{a} & $0.22^{+0.07~(+0.14)}_{-0.07~(-0.13)}$ & $0.17^{+0.06~(+0.11)}_{-0.06~(-0.10)}$ \\
 $i_b$ (\degr)            & $\sin(i)$\tablenotemark{b} & $132.8^{+3.6~(+7.7)}_{-3.1~(-5.7)}$ & $131.0^{+2.9~(+6.4)}_{-2.6~(-4.8)}$ \\
 $\omega_b$ (\degr)       & Uniform(0, 2$\pi$) & $164^{+12~(+22)}_{-13~(-25)}$ & $161^{+12~(+23)}_{-12~(-23)}$ \\
 $\Omega_b$ (\degr)       & Uniform(0, 2$\pi$)\tablenotemark{b} & $170.5^{+4.7~(+10.4)}_{-4.4~(-9.0)}$ & $169.7^{+4.1~(+8.8)}_{-3.6~(-6.9)}$ \\
 $\tau_b$                 & Uniform(0, 1) & $0.403^{+0.035~(+0.074)}_{-0.034~(-0.068)}$ & $0.402^{+0.037~(+0.075)}_{-0.034~(-0.065)}$ \\
 $a_c$ (au)               & LogUniform(1, 100)\tablenotemark{a} & $33.2^{+2.5~(+4.6)}_{-2.3~(-4.0)}$ & $34.3^{+2.2~(+4.6)}_{-1.8~(-3.0)}$ \\
 $e_c$                    & Uniform(0, 1)\tablenotemark{a} & $0.051^{+0.052~(+0.104)}_{-0.035~(-0.049)}$ & $0.037^{+0.041~(+0.088)}_{-0.025~(-0.035)}$ \\
 $i_c$ (\degr)            & $\sin(i)$\tablenotemark{b} & $131.6^{+3.0~(+6.1)}_{-2.9~(-5.2)}$ & $130.5^{+2.5~(+4.9)}_{-2.4~(-5.2)}$ \\
 $\omega_c$ (\degr)       & Uniform(0, 2$\pi$) & $56^{+27~(+60)}_{-24~(-45)}$ & $53^{+29~(+65)}_{-23~(-43)}$ \\
 $\Omega_c$ (\degr)       & Uniform(0, 2$\pi$)\tablenotemark{b} & $161.0^{+4.2~(+8.4)}_{-4.9~(-10.4)}$ & $161.7^{+4.1~(+8.1)}_{-4.3~(-8.8)}$ \\
 $\tau_c$                 & Uniform(0, 1) & $0.51^{+0.09~(+0.19)}_{-0.08~(-0.14)}$ & $0.49^{+0.09~(+0.20)}_{-0.07~(-0.13)}$ \\
 Parallax (mas)         & $\mathcal{N}$(8.8159, 0.0405) & $8.821^{+0.041~(+0.080)}_{-0.042~(-0.081)}$ & $8.821^{+0.041~(+0.084)}_{-0.040~(-0.079)}$ \\
 $M_{b}$ ($M_\textrm{Jup}$)  & Uniform(1, 15) & $7.9^{+4.9~(+6.8)}_{-4.7~(-6.5)}$ & $3.2^{+3.3~(+8.4)}_{-1.6~(-2.1)}$ \\
 $M_{c}$ ($M_\textrm{Jup}$)  & Uniform(1, 15) & $7.8^{+5.0~(+6.9)}_{-4.7~(-6.4)}$ & $7.5^{+4.7~(+7.0)}_{-4.2~(-6.1)}$ \\
 $M_{*}$ ($M_\odot$)  & $\mathcal{N}$(0.88, 0.09) & $0.967^{+0.065~(+0.131)}_{-0.065~(-0.126)}$ & $0.982^{+0.066~(+0.128)}_{-0.066~(-0.130)}$ \\
\enddata
\tablecomments{The orbital parameters used here are defined in \citet{Blunt2020} and are in Jacobi coordinates. For each parameter, the median value of the posterior is listed, with superscript and subscript describing the 68\% credible interval (95\% credible interval in parentheses).}
\tablenotetext{a}{Additional prior on periastron of c is larger than apastron of b}
\tablenotetext{b}{Additional Gaussian prior on the coplanarity of b, c, and the disk}
\end{deluxetable*}

\subsection{Dynamical Constraints}\label{sec:dynamics}

\begin{figure}
    \centering
    \includegraphics[width=0.49\textwidth]{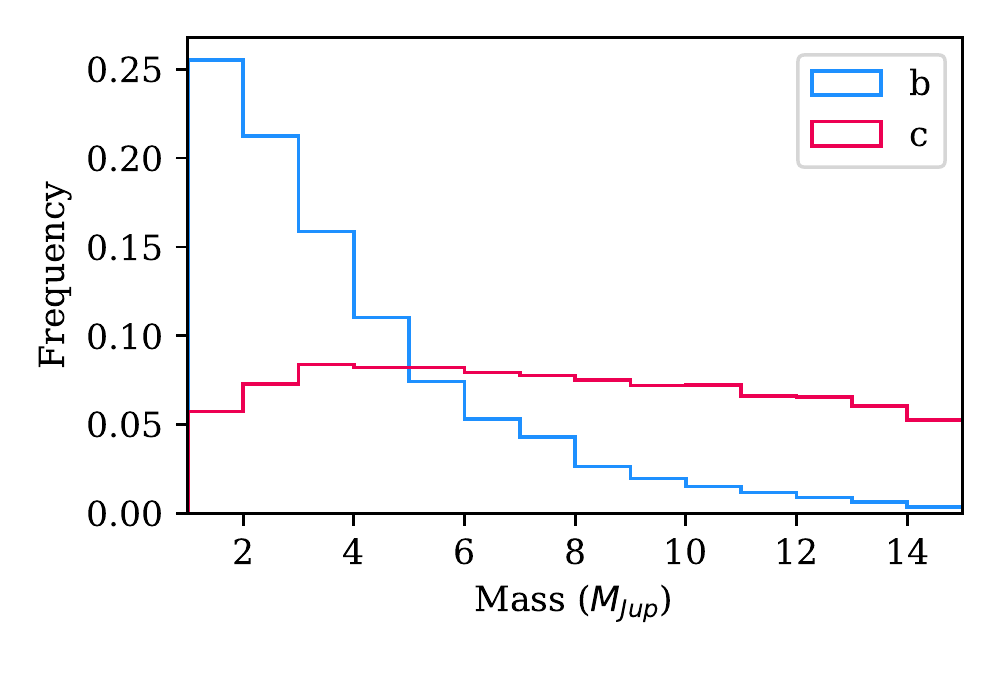}
    \caption{Mass constraints on PDS 70 b and c based on the dynamical stability prior. The mass of PDS 70 c is unconstrained, whereas PDS 70 b disfavors high masses. }
    \label{fig:dynmass}
\end{figure}

The period ratio, slight eccentricity, and masses of the planets bear strong resemblance to the HR 8799 system where at least the innermost two planets (HR 8799 d and e) harbor eccentricities near 0.1, are likely locked in a 2:1 MMR, and have a period ratio slightly larger than 2. \citet{Wang2018} proposed that HR 8799 d and e arrived at this orbital configuration due to resonant migration in the protoplanetary disk: radial migration in MMR excites the planets' eccentricity while eccentricity damping due to the viscous circumstellar disk repelled the planets to period ratios greater than 2.

We investigated whether PDS 70 b and c are in a similar dynamical scenario by searching for dynamically stable orbits. Following the procedure in \citet{Wang2018}, we performed rejection sampling on our posterior of orbital parameters by imposing a stability prior that the orbital configuration is stable for the system's age of 8 Myr, noting our results are not extremely sensitive to the exact choice of age. For each orbit in the posterior, we used the \texttt{REBOUND} $N$-body package with the IAS15 integrator to advance the system backward in time for 8 Myr \citep{Rein2012,Rein2015}. We assigned component masses for all three bodies based on the mass posteriors from our orbit fit. Note that only the mass of the star was constrained by our orbit fit, and that the posterior masses from the two protoplanets were dictated by the prior. We considered a system unstable if the two planets pass within one mutual hill radius of each other ($\sim$3.3 au) or if any planet is ejected to 500 au. During the simulations, we logged the 2:1 resonance angle between the two protoplanets that we define as
\begin{equation}
    \theta_{c:b} = \lambda_b - 2\lambda_c + \varpi_b,
\end{equation}
where $\varpi = \Omega + \omega$ is the longitude of periastron, and $\lambda = \varpi + M$ is the mean longitude ($M$ is the mean anomaly). We used the same algorithm as \citet{Wang2018} to identify where in time series of $\theta_{c:b}$ is it librating or circulating (see their figure 7), and saved the fraction of time the angle was librating over 8 Myr in each simulation.

In these simulations that just account for the gravitational interactions of the three bodies, we found that 41\% of the orbital posterior is dynamically stable and only 3\% of the stable orbits have the two planets in resonance lock (where $\theta_{c:b}$ is librating $>95$\% of the time). In fact, the majority of stable orbits had $\theta_{c:b}$ circulating the entire time, indicating that the planets were not in resonance for any significant period of time in the simulations. The relatively small fraction of stable orbits in resonance lock is likely due to the significant uncertainties in the orbital parameters, with many combinations of orbital parameters lying well outside of any region with MMR can occur. This difficulty in finding resonant orbits has also been seen in HR 8799 \citep{Wang2018}. 

However, we note that we have not included planet-disk interactions and gas drag, which could affect the planets' orbits. It also does not rule out that these planets could in the future migrate into orbital resonance. Thus, we will avoid investigating the detailed dynamical interactions of the system with our simulations alone. Encouragingly, \citet{Bae2019} accounted for these effects and showed that having planets in the approximate orbital configuration of PDS 70 b and c migrate into resonance while accreting from the circumstellar disk would create a circumstellar disk and gap that is consistent with the mm wavelength observations and imply mass accretion rates consistent with the H$\alpha$ luminosities. Furthermore, they predicted that such a migration into resonance would pump the eccentricity of PDS 70 b to $\sim$0.1-0.2, which agrees very well with our inferred eccentricity of $0.17 \pm 0.06$ for PDS 70 b assuming dynamical stability. The current observations are thus consistent with planets being in 2:1 resonance. However, we cannot reject other scenarios that do not require the planets to be in resonance at this time.

We used the dynamical stability prior to place upper limits on the masses of the two planets, and plot the 1D marginalized posterior distribution of their masses in Figure \ref{fig:dynmass}. We found nearly no constraint on the mass of PDS 70 c from enforcing stability, but we found a 95\% upper limit of 10~$M_\textrm{Jup}$ for PDS 70 b, consistent with the masses predicted by \citet{Wang2020}.

The mass of the host star is also constrained by the orbital motion of the planets. Given that the masses of young stars are more difficult to constrain from photometry or spectroscopy alone compared to main-sequence stars, the dynamical mass constraints on the host star is another piece of useful information from the orbit fit. We found a stellar mass of $0.982 \pm {0.066}~M_\odot$ in our dynamically stable solutions. Compared to the dynamical mass estimate of $0.875\pm0.03$\,M$_{\odot}$ from fitting velocity maps of the circumstellar gas \citep{Keppler2019} and the model-dependent mass estimate of $0.88\pm0.02$\,M$_{\odot}$ from the stellar SED fit described in Section \ref{sec:redSpectra}, our dynamical mass estimate from the planetary orbits is systematically high, although it is consistent at the 1.6$\sigma$ level. Due to the short orbital arc, there remains $\sim$10\% uncertainties in our dynamical mass estimate from the orbital motion of the planets, since orbital period, semi-major axis, and stellar mass are degenerate. Because of this, using our dynamical mass posterior as a prior in our stellar SED fits results in no change in the derived stellar spectrum. If we instead fix the stellar mass to 1~M$_{\odot}$ in our SED fits, the $K$-band excess predicted by SED fits drops to 10\%, but the $J$ and $H$ band photometry become 2$\sigma$ discrepant with the model. Extending the orbital coverage with more astrometric monitoring will improve our stellar mass estimate.

\section{Spectral Analysis}\label{sec:sed}

We investigated the nature of the emission from PDS 70 b and c with the additional constraints provided by our GRAVITY $K$-band spectra. We followed the same approach as \citet{Wang2020}, who found that a single blackbody was the best description of the spectral energy distribution of both planets. We investigated whether a blackbody remains the best model for the photospheric emission we observe, or whether more complex models are needed.

We fit the following forward models to the data: a blackbody, the BT-SETTL atmospheric models \citep{Allard2012}, the DRIFT-PHOENIX atmospheric models \citep{Woitke2003, Woitke2004, Helling2006, Helling2008}, and the Exo-REM atmospheric models \cite{Charnay2018}. For all four models, we also considered augmenting each forward model with extinction prescriptions to emulate dust reddening and with a second blackbody element to emulate circumplanetary dust emission. We note that, for the extinction prescriptions, we were agnostic to whether the dust is in the planet's atmosphere, surrounding the planet, or in the circumplanetary or circumstellar disk. Interstellar reddening was found to be negligible \citep{Wang2020}. 

\subsection{PDS 70 b SED Fitting}\label{sec:b_sed_fit}
We used the following literature measurements in the fits along with our GRAVITY $K$-band spectrum: VLT/SPHERE $YJH$ spectrum at R$\sim$30 \citep{Muller2018}, VLT/SPHERE photometry at $H$ and $K$ bands \citep{Muller2018}, and 3-5 $\mu$m photometry from Keck/NIRC2, Gemini/NICI, and VLT/NACO \citep{Muller2018,Wang2020,Stolker2020}. All of the literature photometry we used are listed in Table \ref{table:lit_phot} in Appendix~\ref{sec:lit_data}. We excluded fitting the VLT/SINFONI spectrum from \citet{Christiaens2019APJ} as it disagrees with both the GRAVITY spectrum and the SPHERE photometry at the same wavelengths by being $\sim$30\% brighter. Whether this is astrophysical variability (the SINFONI data was taken $\sim$4 years earlier) or instrumental systematics is uncertain at this point, so we did not consider it here for simplicity. 

With only a tentative water absorption band between the $J$ and $H$ bands, \citet{Wang2020} found that a single blackbody was the most justified model. However, our GRAVITY spectrum shows a dip at the blue end of the $K$ band that is consistent with the water-absorption band seen in substellar atmospheres. To perform this test quantitatively, we performed Bayesian model comparison between the different fits. We fit each model using the same Bayesian framework as \citet{Wang2020}. We used a Gaussian process with the same square exponential kernel to empirically estimate the correlated noise in the SPHERE $YJH$ spectrum when fitting the atmospheric models to the data. The GRAVITY spectrum has its covariance estimated as part of the data reduction and we used this covariance matrix when accounting for its correlated noise in the likelihood.

For the priors, we picked uniform priors in effective temperature ($T_\textrm{eff}$) between 1000 K and 1500 K and uniform priors in effective radius ($R$) between 0.5 and 5 Jupiter radii. For grids with surface gravity (log(g)), metalicity ([M/H]), and carbon to oxygen ratio (C/O), we used uniform priors with the bounds spanned by the edges of the model grids: for BT-SETTL, $3.5 < \log(g) < 5.5$; for DRIFT-PHOENIX, $3.0 < \log(g) < 5.5$ and $-0.3 < [\textrm{M/H}] < 0.3$; for Exo-REM, $3.0 < \log(g) < 4.5$, $-0.5 < [\textrm{M/H}] < 0.5$, and $0.3 < \textrm{C/O} < 0.75$. We used \texttt{pymultinest} to sample the posterior distribution and numerically compute the evidence of each model \citep{Buchner2014}. The median and 95\% credible intervals of each parameter are listed in Table \ref{table:forwardmodels} in the ``Plain Models" section. The evidence allows us to compute the Bayes factor $B$ to test the relative probability of two models:

\begin{equation}
    B \equiv \frac{P(M_1 | D)}{P(M_2 | D)} = \frac{P(D| M_1) P(M_1)}{P(D | M_2) P(M_2)}.
\end{equation}

In this equation, $P$ is the probability of a quantity, $M_1$ and $M_2$ are the two models that are being compared, and $D$ is the data. The left hand side is the relative probability of $M_1$ compared to $M_2$ given the current data. On the right hand side, $P(D | M)$ is the evidence of a given model, and $P(M)$ is the prior probability of a given model.
Assuming equal weight for all models, as we do not think one model is better justified than any of the others, the Bayes factor of two models is equal to the ratio of evidences.
We benchmarked all of the models we considered against the simple blackbody model (i.e., we set it as $M_2$) given it has been the preferred model in previous work \citep{Wang2020,Stolker2020}. We list the values of $B$ for each model relative to the plain blackbody model in the rightmost column of Table \ref{table:forwardmodels}.

\begin{deluxetable*}{c|ccccc|c|cc|c}
\tablecaption{Model fits to SED of PDS 70 b\label{table:forwardmodels}}
\tablehead{
 Model & $T_\textrm{eff}$ (K) & $R$ ($R_{\textrm{Jup}}$) & $\log(g)$ & [M/H] & C/O & $A_V$ (mag) & $a_\textrm{max}$ ($\mu$m) & $\beta_\textrm{dust}$ &  $B$
}
\startdata
\multicolumn{6}{c|}{Plain Models} & & & &  \\
\hline
Blackbody & $1244^{+50}_{-41}$ & $2.47^{+0.24}_{-0.24}$ & - & - & - & - & - & - &  1 \\
BT-SETTL & $1252^{+22}_{-19}$ & $1.97^{+0.05}_{-0.03}$ & $3.51^{+0.04}_{-0.01}$\tablenotemark{a} & - & - & - & - & - &  $5.5 \times 10^{-6}$ \\
DRIFT-PHOENIX & $1384^{+47}_{-48}$ & $1.85^{+0.24}_{-0.16}$ & $5.25^{+0.24}_{-1.43}$ & $-0.15^{+0.36}_{-0.15}$ & - & - & - & - &  96 \\
Exo-REM & $1051^{+17}_{-12}$ & $3.62^{+0.15}_{-0.44}$ & $3.50^{+0.01}_{-0.00}$\tablenotemark{a} & $0.04^{+0.46}_{-0.05}$ & $0.70^{+0.01}_{-0.05}$ & -  & - &  - &  $5.3 \times 10^{-9}$ \\
\hline
\multicolumn{6}{c|}{ISM Extinction} & & & &  \\
\hline
Blackbody & $1352^{+137}_{-109}$ & $2.16^{+0.37}_{-0.30}$ & - & - & - & $2.9^{+3.7}_{-2.7}$ & - & - &  0.69 \\
BT-SETTL & $1418^{+69}_{-106}$ & $1.98^{+0.19}_{-0.14}$ & $3.87^{+0.77}_{-0.36}$ & - & - & $6.1^{+3.3}_{-2.2}$ & - & - &  84 \\
DRIFT-PHOENIX & $1442^{+52}_{-71}$ & $1.85^{+0.25}_{-0.18}$ & $4.62^{+0.84}_{-1.21}$ & $-0.12^{+0.36}_{-0.17}$ & - & $3.0^{+4.7}_{-2.7} $ & - & - &  59 \\
Exo-REM & $1234^{+95}_{-78}$ & $2.88^{+0.44}_{-0.35}$ & $3.89^{+0.42}_{-0.35}$ & $0.01^{+0.44}_{-0.37}$ & $0.61^{+0.12}_{-0.21}$ & $8.3^{+1.5}_{-2.2}$  & - &  - &  46 \\
\hline
\multicolumn{6}{c|}{Power-Law Dust Extinction} & &  &  & \\
\hline
Blackbody & $1273^{+71}_{-55}$ & $2.34^{+0.27}_{-0.29}$ & - & - & - & $2.9^{+6.4}_{-2.8}$ & $0.102^{+0.592}_{-0.078} $ & $-2.4^{+2.3}_{-2.8}$ & 0.14 \\
BT-SETTL & $1402^{+71}_{-89}$ & $2.53^{+0.74}_{-0.80}$ & $3.89^{+0.53}_{-0.35}$ & - & - & $3.5^{+4.3}_{-2.6}$ & $2.41^{+6.26}_{-2.13}$ & $-3.9^{+3.1}_{-1.1}$ &  3.8 \\
DRIFT-PHOENIX & $1413^{+54}_{-59}$ & $1.80^{+0.27}_{-0.18}$ & $5.04^{+0.43}_{-1.58}$ & $-0.07^{+0.33}_{-0.21}$ & - & $3.1^{+6.3}_{-2.9} $ & $0.097^{+0.369}_{-0.073}$ & $-2.4^{+2.2}_{-2.7}$ &  20 \\
Exo-REM & $1223^{+105}_{-79}$ & $2.81^{+0.52}_{-0.45}$ & $3.84^{+0.38}_{-0.31}$ & $0.00^{+0.43}_{-0.39}$ & $0.61^{+0.12}_{-0.17}$ & $4.7^{+4.1}_{-2.9}$  & $0.70^{+6.49}_{-0.28}$ &  $-4.1^{+3.5}_{-1.1}$ &  1.1 \\
\hline
\hline
\multicolumn{6}{c|}{2nd Blackbody} & & $T_2$ (K) & $R_2$ ($R_{\textrm{Jup}}$) & \\
\hline
Blackbody & $1252^{+48}_{-40}$ & $2.42^{+0.24}_{-0.23}$ & - & - & - &  -  & $351^{+190}_{-225}$ & $24.5^{+23.3}_{-19.7}$ &  0.53 \\
BT-SETTL & $1266^{+68}_{-81}$ & $1.34^{+0.36}_{-0.42}$ & $3.88^{+0.60}_{-0.35}$ & - & - &  -  & $1037^{+107}_{-182}$ & $3.1^{+1.3}_{-0.7}$ &  15 \\
DRIFT-PHOENIX & $1393^{+54}_{-45}$ & $1.81^{+0.21}_{-0.19}$ & $5.30^{+0.19}_{-1.28}$ & $-0.11^{+0.38}_{-0.18}$ & - &  -  & $301^{+147}_{-189}$ & $28.4^{+20.1}_{-24.2}$ &  38 \\
Exo-REM & $1121^{+80}_{-59}$ & $1.44^{+0.59}_{-0.39}$ & $4.16^{+0.31}_{-0.50}$ & $0.14^{+0.32}_{-0.47}$ & $0.54^{+0.16}_{-0.22}$ &  -  & $1078^{+59}_{-83}$ & $3.1^{+0.6}_{-0.4}$ &  0.44 \\
\hline
\multicolumn{6}{c|}{ISM Extinction + 2nd Blackbody} & &  &  & \\
\hline
Blackbody & $1367^{+123}_{-119}$ & $2.11^{+0.38}_{-0.27}$ & - & - & - &  $3.1^{+3.2}_{-2.9} $  & $369^{+158}_{-241}$ & $26.5^{+21.7}_{-21.4}$ &  0.34 \\
BT-SETTL & $1392^{+82}_{-82}$ & $1.96^{+0.20}_{-0.17}$ & $3.83^{+0.69}_{-0.32}$ & - & - &  $5.4^{+3.4}_{-1.9}$  & $407^{+265}_{-128}$ & $26.6^{+21.5}_{-21.6}$ &  147 \\
DRIFT-PHOENIX & $1449^{+47}_{-54}$ & $1.84^{+0.26}_{-0.18}$ & $5.11^{+0.38}_{-1.67}$ & $-0.11^{+0.33}_{-0.18}$ & - &  $3.0^{+4.5}_{-2.2}$  & $384^{+323}_{-263}$ & $3.8^{+4.8}_{-1.9}$ &  2.2 \\
Exo-REM & $1235^{+91}_{-76}$ & $2.86^{+0.34}_{-0.33}$ & $3.89^{+0.36}_{-0.36}$ & $0.03^{+0.43}_{-0.32}$ & $0.60^{+0.13}_{-0.21}$ &  $8.0^{+1.7}_{-2.2}$  & $318^{+207}_{-202}$ & $24.7^{+23.4}_{-20.2}$ &  14 \\
\hline
\hline
\multicolumn{6}{c|}{Accreting CPD} & & $M_p \dot{M}$ ($M_\textrm{Jup}^2$/yr) & $R_{in}$ ($R_{\textrm{Jup}}$) & \\
\hline
Blackbody & $1257^{+50}_{-43}$ & $2.38^{+0.24}_{-0.26}$ & - & - & - &  -  & $3.3^{+6.4}_{-2.1} \times 10^{-7}$ & $2.7^{+1.1}_{-1.2}$ &  0.24 \\
BT-SETTL & $1259^{+126}_{-26}$ & $1.68^{+0.12}_{-0.25}$ & $3.55^{+0.18}_{-0.05}$\tablenotemark{a} & - & - &  -  & $2.9^{+11.5}_{-0.8}  \times 10^{-7}$ & $1.1^{+2.2}_{-0.1}$ &  0.55 \\
DRIFT-PHOENIX & $1409^{+45}_{-47}$ & $1.76^{+0.23}_{-0.17}$ & $5.26^{+0.22}_{-1.33}$ & $0.02^{+0.26}_{-0.30}$ & - &  -  & $3.2^{+6.0}_{-2.1} \times 10^{-7}$ & $2.9^{+1.1}_{-1.2}$ &  14 \\
Exo-REM & $1076^{+77}_{-43}$ & $2.59^{+0.38}_{-0.42}$ & $3.52^{+0.11}_{-0.02}$\tablenotemark{a} & $0.26^{+0.22}_{-0.29}$ & $0.66^{+0.05}_{-0.16}$ &  -  & $2.8^{+2.7}_{-0.7} \times 10^{-7}$ & $1.1^{+0.2}_{-0.1}$ &  $2.8 \times 10^{-5}$ 
\enddata
\tablecomments{For each parameter, a 95\% credible interval centered about the median is reported. The superscript and subscript denote the upper and lower bounds of that range. }
\tablenotetext{a}{Mode of posterior reached edge of model grid}
\end{deluxetable*}

Given the accreting nature of PDS 70 b, it is possible that the planet is extincted by accreting materials or by circumplanetary and circumstellar dust. Indeed previous atmosphere modeling indicated that both planets should be shrouded by its dust \citep{Wang2020, Stolker2020}. The emission we observed could be a planetary atmosphere attenuated by obscuring dust. Although it is not entirely accurate, we first considered a simple interstellar medium (ISM) extinction law. An ISM extinction law has been shown to be an adequate approximation of stars shrouded by their circumstellar disks, so it is not unreasonable \citep{Looper2010a, Looper2010b}. We used an extinction law derived for stars attenuated by the interstellar medium in the near-infrared by \citet{Wang2019}. This extinction law follows the form of
\begin{equation}
    A_\lambda = A_V \left( \frac{\lambda}{\lambda_V} \right)^{-\beta},
\end{equation}
where $A_\lambda$ is the magnitudes of extinction at a wavelength $\lambda$, $A_V$ is the extinction in the $V$ band with center wavelength $\lambda_V = 0.55~\mu\textrm{m}$, and $\beta$ is the power-law index of 2.07 derived by \citet{Wang2019}. As we fixed the power-law index, $A_V$ is the only new free variable introduced. We placed a uniform prior on $A_V$ between 0 and 10 mags. We repeated the fit of the four model grids, but now with the model flux attenuated by
\begin{equation}
    F_\textrm{obs} = 10^{-A_\lambda/2.5} F_\textrm{emit},
\end{equation}
where $F_{obs}$ is the observed flux and $F_{emit}$ is the original flux from the model grids. We recorded the best-fit parameters and $B$ relative to the plain blackbody model with no extinction in Table \ref{table:forwardmodels} under ``ISM Extinction."

Given that the ISM extinction law may not be fully representative of accreting dust in a circumplanetary environment where we expect grain growth \citep[e.g.,][]{Birnstiel2012, Kataoka2013, Piso2015}, a more general case where the dust particles follow a variable power law in grain sizes may better describe the data. Such dust extinction prescriptions have been shown to fit dusty free-floating brown dwarfs \citep{Marocco2014, Hiranaka2016} and directly imaged companions \citep{bonnefoy2016,Delorme2017}. Thus, we considered replacing the ISM extinction law with a power-law dust extinction prescription. We assumed MgSiO$_3$ dust with particle size distribution $n \propto a^{\beta_\textrm{dust}}$ with $a$ being the radius of the dust, and $\beta_\textrm{dust}$ being the power-law exponent. We set a minimum dust radius of 1 nm, and vary the maximum dust radius $a_\textrm{max}$, as the minimum grain size does not significantly impact the spectrum. For a given $a_\textrm{max}$ and $\beta_\textrm{dust}$, we computed the extinction cross section ($\sigma_\textrm{dust}$) of the dust as a function of wavelength with \texttt{PyMieScatt} \citep{sumlin2018} by using the refractive indices from \citet{scott1996} and \citet{jaeger1998}. To relate the cross-sectional area to an amount of attenuated flux per wavelength, we used the relation
\begin{equation}
    F_\textrm{obs} = e^{-\tau_\textrm{dust}} F_\textrm{emit} = 10^{-A_V/2.5} e^{-\frac{\sigma_\textrm{dust}}{\sigma_\textrm{dust,V}}} F_\textrm{emit},
\end{equation}
where $\sigma_\textrm{dust,V}$ is the cross-sectional absorption area averaged across the $V$ band, and $\tau_\textrm{dust}$ is the optical depth of the dust. Our uniform prior on $a_\textrm{max}$ was between 0.01 and 10 $\mu$m and and our uniform prior on $\beta_\textrm{dust}$ was between -10 and 0. Our prior on $A_V$ remained between 0 and 10 mags.

We also considered augmenting the forward models with circumplanetary disk (CPD) models. We first used a simple blackbody component to model the CPD as has been considered in the past work \citep{Wang2020, Stolker2020}. CPD models have indicated that the bulk of the thermal emission from a circumplanetary disk would come from the inner edge of the disk \citep{Zhu2015, Szulagyi2019}. For the second blackbody, we adopted priors for the temperature of the second blackbody component ($T_2$ and $R_2$ respectively) that are motivated by these modeling studies. The $T_2$ prior was a uniform prior between 100~K and $T_\textrm{eff}$, the effective temperature of the first component. The $R_2$ prior was a uniform prior between $R$, the effective radius of the first component, and 50 $R_\textrm{Jup}$. These priors are not uniform in $T_2$ and $R_2$, but if we marginalized over $T_\textrm{eff}$ and $R$, we get priors that only weakly favor lower temperatures and larger radii.

Since extinction of the planetary atmosphere may play a significant role, we considered the case of an extincted atmosphere model plus a second blackbody component, similar to what was done in \citet{Christiaens2019APJ}. This case attempts to model circumplanetary dust absorbing the light from the protoplanet and reradiating it away at longer wavelengths. We used the simple ISM extinction law, as it has fewer free parameters, even though we note that the slope may not be perfectly accurate for circumplanetary dust. We only applied the extinction to the atmospheric model and not the second blackbody component. We used the same priors on $A_V$, $T_2$, and $R_2$ as previously.

Last, we considered using the more sophisticated accreting CPD model from \citet{Zhu2015} that model the emission from a CPD with density and temperature gradients and account for molecular and atomic opacities. The resulting spectra are parameterized by the product of the planet's mass and its mass-accretion rate ($M_p \dot{M}$) and the inner edge of the CPD ($R_{in}$). The spectra are only weakly sensitive to the outer disk edge. \citet{Zhu2015} produced models with the outer radius being 50 and 1000 $R_{in}$, and we marginalized over the two outer radii in our SED fits, as there was no statistically significant difference. 

\begin{figure*}
    \centering
    \includegraphics[width=\textwidth]{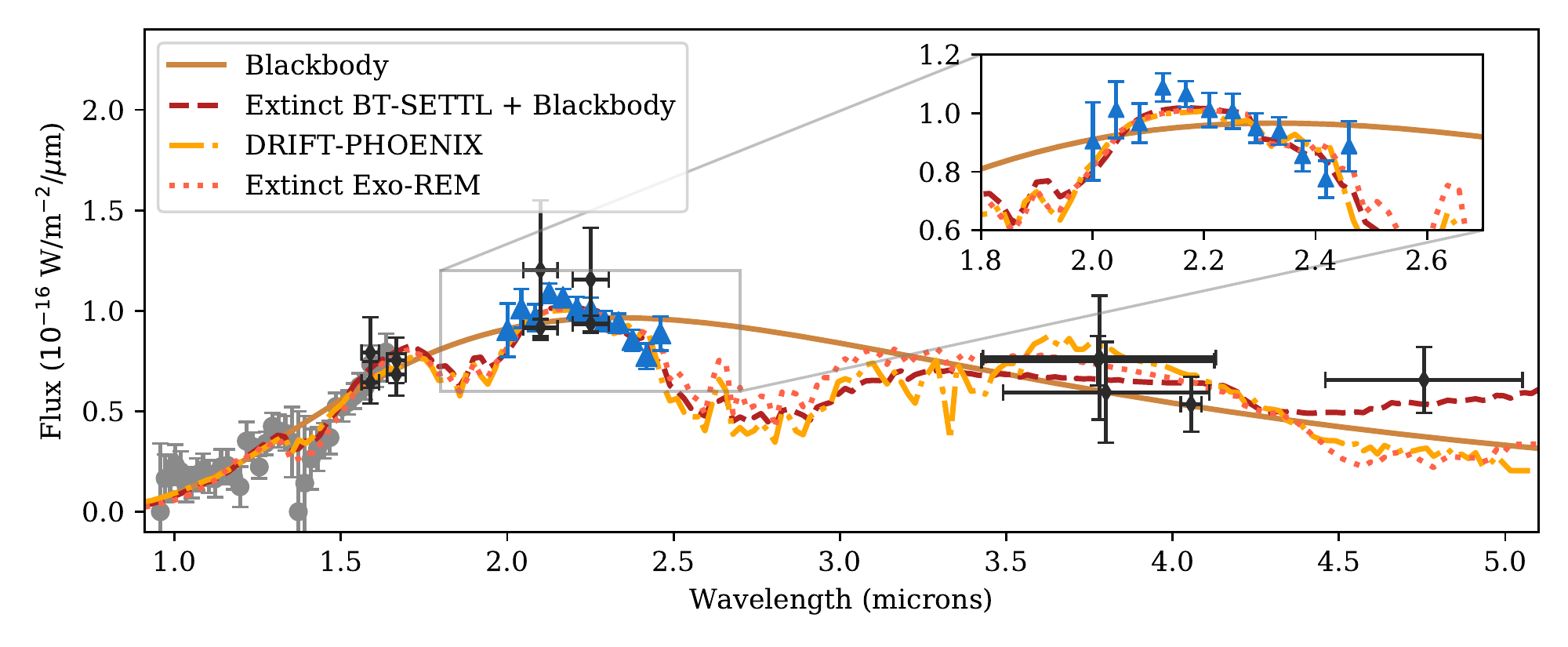}
    \caption{Spectral energy distribution data and models for PDS 70 b. For each of the four forward model grids, we plot the best-fit model from the modification case with the highest Bayes factor. The data are also overplotted. The GRAVITY spectrum (in blue) is binned with each point representing the weighted mean of 19 spectral channels and the error bar is the 1$\sigma$ weighted error of the binned flux (note that the fits were still done on the unbinned data). The SPHERE IFS spectrum is the gray, and the literature photomety is in black. The inset plot zooms in on the $K$-band region, plotting the models and only the GRAVITY data for comparison.}
    \label{fig:bsedfit}
\end{figure*}

In all, we tried six different modifications to the four forward models, resulting in 24 models.  We plotted the best-fit model for the model modification with the highest Bayes factor for each of the four atmospheric forward models in Figure \ref{fig:bsedfit} and listed all the results in Table \ref{table:forwardmodels}. We discussed the model selection and implications further in Section \ref{sec:sed_model_comp}.

\subsection{PDS 70 c SED Fit}\label{sec:c_sed_fit}
In addition to the GRAVITY $K$-band spectrum and the re-extracted SPHERE IFS $YJH$ spectrum, we also used the $K$-band photometry from \citet{Mesa2019} and $L$-band photometry from \citet{Wang2020}. We listed the exact numbers for the literature photometry that we used in Table \ref{table:lit_phot} in Appendix \ref{sec:lit_data}. In this systematic exploration of models, we did not fit the 855 $\mu$m continuum emission coming from the location of PDS 70 c \citep{Isella2019}, as the Exo-REM and accreting CPD grids did not extend to those wavelengths, and to primarily focus on the 1-5 $\mu$m SED where the bulk of the planetary emission should be (see the end of the section for more fits with this data point).  

We used the same four base forward models. We modified the prior for $T_\textrm{eff}$ of the blackbody models to be between 700 to 1200~K instead, as the previous prior range for PDS 70 b was too high. We did not modify the $T_\textrm{eff}$ for the other forward models because they did not go below 1000~K. For all four models, the range of $T_\textrm{eff}$ remained 500~K, so the impact of a different $T_\textrm{eff}$ prior on the evidence of the blackbody models should be negligible.

We also used the same extinction and CPD modifications as for PDS 70 b. The only change was changing the prior limits for $A_V$ to be between 0 and 20 mags instead of 0 and 10 mags, as preliminary analysis indicated the extinction could be greater than 10 mags. Increasing the prior range on $A_V$ may decrease the evidence of the models with extinction slightly, but we accepted this in order to have a more flexible extinction prescription.

\begin{figure*}
    \centering
    \includegraphics[width=\textwidth]{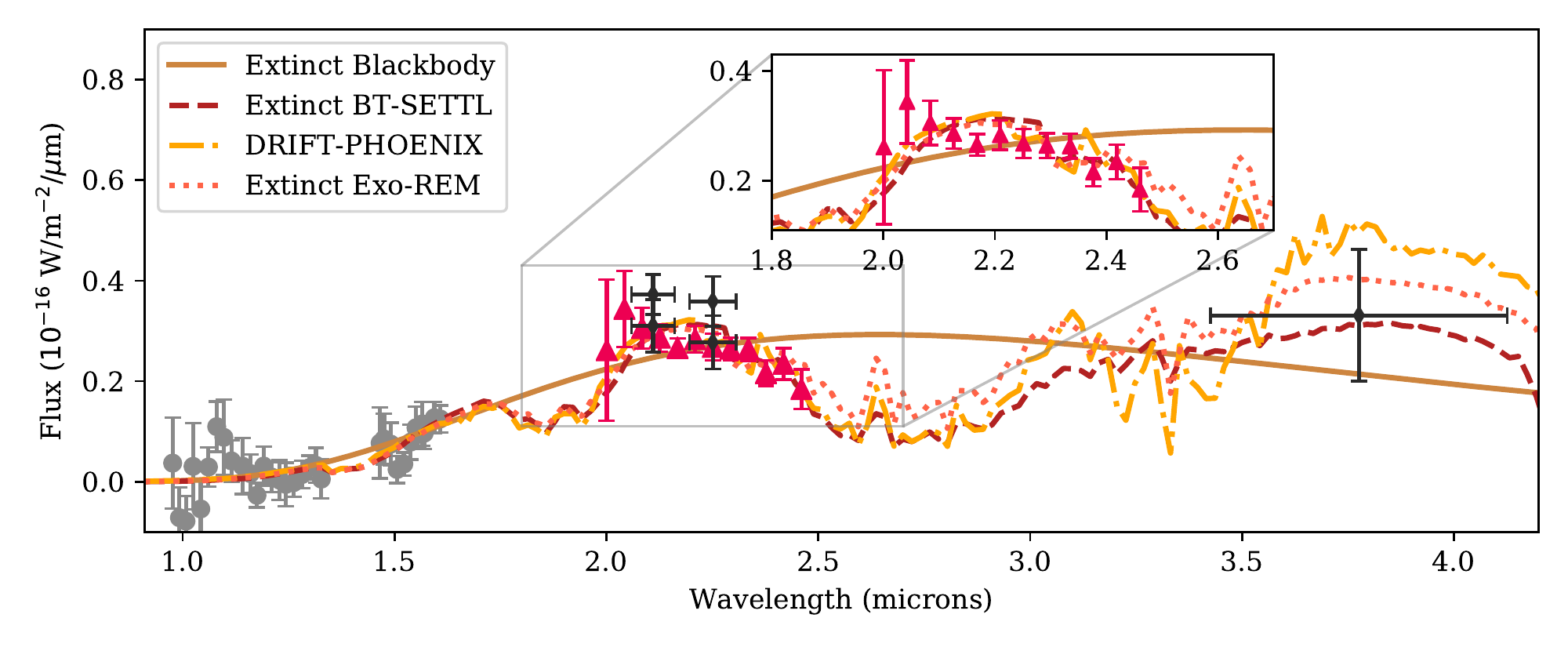}
    \caption{Same plot as Figure \ref{fig:bsedfit} except for PDS 70 c. To make it clearer to see by eye, both GRAVITY datasets have been binned together into a single spectrum (in red) using a weighted mean for each bin.}
    \label{fig:csedfit}
\end{figure*}

The 95\% CI centered about the median of each parameter of each model fit along with the Bayes factor of each model relative to the single Blackbody model with no modifications are listed in Table \ref{table:c_forwardmodels}. The best-fit spectrum of the model modification with the highest Bayes factor for each of the four base forward models are plotted in Figure \ref{fig:csedfit}.

Given that \citet{Isella2019} used the 855 $\mu$m detection to demonstrate the existence of a CPD disk around PDS 70 c, we ran a few fits including this photometric point to verify this conclusion and characterize the CPD. As baseline models, we repeated the single and two blackbody fits with this longer wavelength measurement. From the fits above to the 1-5 $\mu$m data, we found that the model with the most support from the data was the plain DRIFT-PHOENIX, and that augmenting it with a cooler blackbody component was acceptable (see Table \ref{table:c_forwardmodels} and Section \ref{sec:sed_model_comp}). We thus also refit the plain DRIFT-PHOENIX model and the DRIFT-PHOENIX model supplemented with a cooler blackbody component. In the fits, we extended the upper limit on the prior for $R_2$ to 5000~$R_{Jup}$ (2.4~au) and the lower limit of $T_2$ to 10~K based on the results from \citep{Isella2019} and \citet{Wang2020} that point to a very large and cold CPD. Since the data used in the fit changed, we avoided direct model comparisons between these fits and the fits to only the 1-5 $\mu$m data, and only compared these four models among themselves. We define $B_{855}$ as the Bayes factor between one of these models and the plain blackbody fit that includes the 855 $\mu$m data point. We list the results of the model fits in Table \ref{table:c_alma_forwardmodels}.

\begin{deluxetable*}{c|ccccc|c|cc|c}
\tablecaption{Model fits to SED of PDS 70 c\label{table:c_forwardmodels}}
\tablehead{
 Model & $T_\textrm{eff}$ (K) & $R$ ($R_{\textrm{Jup}}$) & $\log(g)$ & [M/H] & C/O & $A_V$ (mag) & $a_\textrm{max}$ ($\mu$m) & $\beta_\textrm{dust}$ &  $B$
}
\startdata
\multicolumn{6}{c|}{Plain Models} & & & &  \\
\hline
Blackbody & $1005^{+65}_{-58}$ & $2.39^{+0.53}_{-0.42}$ & - & - & - & - & - & - &  1 \\
BT-SETTL & $1155^{+37}_{-8}$ & $1.27^{+0.05}_{-0.12}$ & $3.50^{+0.01}_{-0.00}$\tablenotemark{a} & - & - & - & - & - &  $3.4 \times 10^{-11}$ \\
DRIFT-PHOENIX & $1054^{+60}_{-48}$ & $1.98^{+0.39}_{-0.31}$ & $4.54^{+0.61}_{-0.85}$ & $-0.22^{+0.46}_{-0.07}$ & - & - & - & - &  $6.2 \times 10^{7}$ \\
Exo-REM & $1024^{+30}_{-22}$ & $1.89^{+0.26}_{-0.21}$ & $3.50^{+0.02}_{-0.00}$\tablenotemark{a} & $0.47^{+0.03}_{-0.15}$\tablenotemark{a} & $0.65^{+0.02}_{-0.02}$ & -  & - &  - &  $3.1 \times 10^{-17}$ \\
\hline
\multicolumn{6}{c|}{ISM Extinction} & & & &  \\
\hline
Blackbody & $1401^{+92}_{-289}$ & $1.40^{+0.65}_{-0.16}$ & - & - & - & $14.1^{+3.8}_{-8.7}$ & - & - &  19 \\
BT-SETTL & $1289^{+99}_{-104}$ & $1.60^{+0.30}_{-0.19}$ & $4.10^{+0.54}_{-0.54}$ & - & - & $18.1^{+1.8}_{-4.0}$ & - & - &  $3.6 \times 10^{6}$ \\
DRIFT-PHOENIX & $1086^{+128}_{-68}$ & $1.97^{+0.44}_{-0.49}$ & $5.05^{+0.41}_{-0.81}$ & $-0.11^{+0.37}_{-0.17}$ & - & $5.3^{+5.1}_{-4.4} $ & - & - &  $5.5 \times 10^{7}$ \\
Exo-REM & $1112^{+133}_{-98}$ & $2.42^{+0.66}_{-0.54}$ & $3.87^{+0.48}_{-0.30}$ & $0.01^{+0.44}_{-0.42}$ & $0.50^{+0.20}_{-0.19}$ & $18.6^{+1.3}_{-3.2}$  & - &  - &  $2.2 \times 10^{6}$ \\
\hline
\multicolumn{6}{c|}{Power-Law Dust Extinction} & &  &  & \\
\hline
Blackbody & $1338^{+150}_{-279}$ & $1.23^{+0.88}_{-0.24}$ & - & - & - & $16.0^{+3.8}_{-8.5}$ & $0.31^{+0.24}_{-0.24}$ & $-2.8^{+2.4}_{-2.3}$ &  13 \\
BT-SETTL & $1327^{+148}_{-115}$ & $1.24^{+0.51}_{-0.17}$ & $4.38^{+0.84}_{-0.83}$ & - & - & $15.2^{+4.5}_{-8.5}$ & $0.41^{+2.79}_{-0.12}$ & $-2.5^{+2.3}_{-2.6}$ &  $4.3 \times 10^{5}$ \\
DRIFT-PHOENIX & $1074^{+78}_{-64}$ & $1.92^{+0.50}_{-0.42}$ & $4.73^{+0.66}_{-0.72}$ & $-0.14^{+0.39}_{-0.15}$ & - & $11.3^{+8.0}_{-10.3} $ & $0.089^{+0.595}_{-0.070}$ & $-2.5^{+2.3}_{-3.2}$ &  $2.1 \times 10^{7}$ \\
Exo-REM & $1118^{+130}_{-104}$ & $1.96^{+0.55}_{-0.50}$ & $3.92^{+0.50}_{-0.35}$ & $-0.15^{+0.57}_{-0.33}$ & $0.47^{+0.22}_{-0.15}$ & $16.3^{+3.4}_{-8.1}$  & $0.37^{+0.42}_{-0.08}$ &  $-2.0^{+1.9}_{-2.8}$ &  $2.9 \times 10^{5}$ \\
\hline
\hline
\multicolumn{6}{c|}{2nd Blackbody} & & $T_2$ (K) & $R_2$ ($R_{\textrm{Jup}}$) & \\
\hline
Blackbody & $1024^{+51}_{-22}$ & $2.25^{+0.18}_{-0.31}$ & - & - & - &  -  & $268^{+190}_{-153}$ & $21.9^{+25.5}_{-18.5}$ &  0.12 \\
BT-SETTL & $1152^{+47}_{-33}$ & $0.80^{+0.19}_{-0.18}$ & $3.52^{+0.09}_{-0.02}$\tablenotemark{a} & - & - &  -  & $811^{+72}_{-104}$ & $4.0^{+2.4}_{-0.9}$ &  $4.1 \times 10^{-3}$ \\
DRIFT-PHOENIX & $1054^{+63}_{-48}$ & $1.96^{+0.30}_{-0.30}$ & $4.46^{+0.48}_{-0.76}$ & $-0.19^{+0.41}_{-0.10}$ & - &  -  & $269^{+238}_{-154}$ & $22.4^{+24.5}_{-18.9}$ &  $1.6 \times 10^{7}$ \\
Exo-REM & $1064^{+68}_{-54}$ & $0.62^{+0.24}_{-0.11}$ & $3.94^{+0.38}_{-0.37}$ & $0.22^{+0.25}_{-0.49}$ & $0.53^{+0.19}_{-0.21}$ &  -  & $878^{+54}_{-66}$ & $3.5^{+1.1}_{-0.7}$ &  $3.1 \times 10^{-5}$ \\
\hline
\multicolumn{6}{c|}{ISM Extinction + 2nd Blackbody} & &  &  & \\
\hline
Blackbody & $1186^{+122}_{-155}$ & $1.80^{+0.59}_{-0.38}$ & - & - & - &  $7.8^{+2.1}_{-6.2} $  & $265^{+204}_{-152}$ & $23.4^{+24.1}_{-20.2}$ &  1.3 \\
BT-SETTL & $1291^{+101}_{-91}$ & $1.59^{+0.23}_{-0.20}$ & $4.11^{+0.53}_{-0.54}$ & - & - &  $17.8^{+2.0}_{-4.3}$  & $278^{+238}_{-167}$ & $21.4^{+26.2}_{-18.3}$ & $8.3 \times 10^{5}$ \\
DRIFT-PHOENIX & $1088^{+121}_{-65}$ & $1.90^{+0.37}_{-0.42}$ & $4.91^{+0.46}_{-0.73}$ & $-0.02^{+0.29}_{-0.25}$ & - &  $5.5^{+4.7}_{-4.8}$  & $252^{+207}_{-136}$ & $24.8^{+23.1}_{-20.8}$ &  $9.6 \times 10^{6}$ \\
Exo-REM & $1123^{+117}_{-104}$ & $2.36^{+0.67}_{-0.51}$ & $3.92^{+0.44}_{-0.34}$ & $0.02^{+0.43}_{-0.34}$ & $0.49^{+0.20}_{-0.17}$ &  $18.5^{+1.4}_{-3.2}$  & $261^{+205}_{-149}$ & $21.6^{+26.6}_{-17.0}$ &  $5.4 \times 10^{5}$ \\
\hline
\hline
\multicolumn{6}{c|}{Accreting CPD} & & $M_p \dot{M}$ ($M_\textrm{Jup}^2$/yr) & $R_{in}$ ($R_{\textrm{Jup}}$) & \\
\hline
Blackbody & $1026^{+56}_{-24}$ & $2.22^{+0.20}_{-0.35}$ & - & - & - &  -  & $2.2^{+6.6}_{-1.1}  \times 10^{-7}$ & $3.1^{+0.8}_{-1.4}$ &  0.038 \\
BT-SETTL & $1153^{+50}_{-27}$ & $0.81^{+0.17}_{-0.22}$ & $3.52^{+0.09}_{-0.02}$\tablenotemark{a} & - & - &  -  & $3.1^{+3.4}_{-1.1}  \times 10^{-7}$ & $1.3^{+0.13}_{-0.15}$ &  $1.8 \times 10^{-4}$ \\
DRIFT-PHOENIX & $1059^{+57}_{-53}$ & $1.89^{+0.28}_{-0.27}$ & $4.40^{+0.53}_{-0.83}$ & $-0.18^{+0.42}_{-0.12}$ & - &  -  & $2.6^{+6.6}_{-1.5}  \times 10^{-7}$ & $3.0^{+0.9}_{-1.3}$ &  $3.4 \times 10^{6}$ \\
Exo-REM & $1051^{+81}_{-47}$ & $0.63^{+0.25}_{-0.12}$ & $3.79^{+0.37}_{-0.27}$ & $0.29^{+0.20}_{-0.-54}$ & $0.55^{+0.18}_{-0.23}$ &  -  & $2.7^{+11.2}_{-0.7}  \times 10^{-7}$ & $1.2^{+2.2}_{-0.2}$ &  $3.2 \times 10^{-4}$ 
\enddata
\tablecomments{For each parameter, a 95\% credible interval centered about the median is reported. The superscript and subscript denote the upper and lower bounds of that range. }
\tablenotetext{a}{Mode of posterior reached edge of model grid}
\end{deluxetable*}

\begin{deluxetable*}{c|cccc|cc|c}
\tablecaption{Model fits to SED of PDS 70 c including 855 $\mu$m photometry \label{table:c_alma_forwardmodels}}
\tablehead{
 Model & $T_\textrm{eff}$ (K) & $R$ ($R_{\textrm{Jup}}$) & $\log(g)$ & [M/H] &  $T_2$ (K) & $R_2$ ($R_{\textrm{Jup}}$) &  $B_{855}$
}
\startdata
\multicolumn{5}{c|}{Plain Models} & & &  \\
\hline
Blackbody & $1005^{+55}_{-57}$ & $2.39^{+0.52}_{-0.37}$ & - & - & - & - &  1 \\
DRIFT-PHOENIX & $1051^{+62}_{-44}$ & $1.99^{+0.33}_{-0.32}$ & $4.53^{+0.57}_{-0.91}$ & $-0.21^{+0.40}_{-0.08}$ & - & - &  $7.0 \times 10^{7}$ \\
\hline
\multicolumn{5}{c|}{2nd Blackbody} & & & \\
\hline
Blackbody & $1021^{+49}_{-20}$ & $2.27^{+0.17}_{-0.31}$ & - &  -  & $119^{+150}_{-105}$ & $430^{+1198}_{-168}$ &  $1.7 \times 10^{4}$ \\
DRIFT-PHOENIX & $1055^{+58}_{-48}$ & $1.96^{+0.28}_{-0.29}$ & $4.46^{+0.45}_{-0.79}$ & $-0.19^{+0.41}_{-0.10}$ & $125^{+132}_{-94}$ & $419^{+518}_{-151}$ &  $1.7 \times 10^{12}$ \\
\enddata
\tablecomments{For each parameter, a 95\% credible interval centered about the median is reported. The superscript and subscript denote the upper and lower bounds of that range. }
\end{deluxetable*}

\subsection{Model Comparison} \label{sec:sed_model_comp}
For the purposes of model selection, we denoted any model within a Bayes factor of 100 of the best-fitting model (i.e., highest Bayes factor) to be ``adequate." The relative probability of adequate models are $> 1$\% compared to the best fitting model, which we considered good enough to not be excluded. First, we discuss the fits that only consider the 1-5 $\mu$m data. For PDS 70 b, we found that the BT-SETTL model modified with both extinction and a second blackbody component has the most support from the data. For PDS 70 c, the plain DRIFT-PHOENIX model has the highest Bayes factor by being able to fit the data the best without unnecessary free parameters. Thus we considered models with $B > 1.5$ and $B > 6 \times 10^{5}$ to be adequate for PDS 70 b and c, respectively. 

Based on the Bayes factors, the new GRAVITY $K$-band spectra are able to reject the pure blackbody model for the photosphere of both protoplanets in favor of the three planetary atmosphere models. In particular, the falling slopes in both the short and long wavelength ends of the $K$ band are incompatible with blackbody predictions (see inset of Figure \ref{fig:bsedfit} and Figure \ref{fig:csedfit}), and require opacity sources such as water, molecular hydrogen, and carbon monoxide absorption to create the observed slopes in the GRAVITY spectra. For PDS 70 b, this corroborates the tentative 1.4 $\mu$m water absorption feature seen in the SPHERE IFS data \citep{Muller2018}. The difference in Bayes factors is far steeper for PDS 70 c. This appears to be due to the fact that the slope of the GRAVITY K-band spectrum for PDS 70 c is in much starker disagreement with the predictions made by the blackbody model. Thus, we will mainly focus on the three planetary atmosphere models, as all three are adequate fits given the appropriate modifications.

The plain DRIFT-PHOENIX model, in addition to being the most favored model for PDS 70 c, is the model with the second highest support for PDS 70 b.
Adding modifications to the DRIFT-PHOENIX model did not improve the fit, resulting in lower Bayes factors. This can also be seen in the range of $A_V$, $T_2$, and $R_2$ parameters derived in the fits with modifications. The ranges of these parameters are typically consistent with the lower bounds of the priors for these parameters, implying they are minimally altering the DRIFT-PHOENIX spectrum.

The BT-SETTL and Exo-REM models, on the other hand, are poor fits to the data without modifications, with a Bayes factor orders of magnitude worse than both the plain DRIFT-PHOENIX and blackbody models. However, adding some sort of extinction to change the overall 1-4 $\mu$m slope drastically improved their fit, pulling their Bayes factor to within a factor of 100 of the best fitting model. The ISM extinction amplitude of $3.9 < A_V < 9.4$ mag for BT-SETTL fits to PDS 70 b corresponds to an $0.23 < A_K < 0.54$ mag, which is similar to the extinction values found for dusty brown dwarfs \citep{Marocco2014, Delorme2017}. 

Switching from ISM extinction with one free parameter to a variable power-law dust extinction with three free parameters caused drops in the Bayes factor in nearly all cases, except for the Exo-REM model of PDS 70 c (although this model's $B$ was too low to be considered adequate). We do not think this implies that we are seeing extinction from ISM-like grains, but rather that the current data are insufficient to constrain more-flexible extinction models. In all cases, we ruled out extreme size distributions with $\beta_\textrm{dust} < -5$ that are dominated solely by small particles. While the maximum dust size ($a_\textrm{max}$) is relatively unconstrained for PDS 70 b, most of our fits ruled out dust particles larger than about 1 $\mu$m for PDS 70 c. However, we note that only the DRIFT-PHOENIX model with power-law dust extinction has an adequate $B$ for PDS 70 c, so it is unclear how robust this conclusion is. Rather, we are worried that the free parameters in the model are compensating for other model deficiencies. Overall, the lack of improvement in the $B$ indicates the current data is unable to characterize the properties of any obscuring dust.

The DRIFT-PHOENIX and Exo-REM models have free parameters to describe the composition of the atmosphere ([M/H] and C/O). In all the adequate fits to the data, these parameters are essentially unconstrained (e.g., [M/H] spans the whole prior range for acceptable DRIFT-PHOENIX models of PDS 70 b). There are a few edge cases that are excluded (e.g., C/O $< 0.4$ is excluded for adequate Exo-REM models of PDS 70 b), but we take such constraints with caution as atmosphere models can spuriously constrain C/O when there are other inaccuracies in the model (e.g., the plain Exo-REM fits to both planets have the smallest uncertainties on C/O, but the lowest $B$ of all models).

For all three planetary atmosphere models, the implied masses based on the retrieved $\log(g)$ and radii generally favor masses $> 10$~$M_\textrm{Jup}$. However, our priors are biased to high masses as the model grids generally do not go down to a sufficiently low surface gravity for the $\sim$2 $R_\textrm{Jup}$ effective radii we measured: a 1 $M_\textrm{Jup}$ and 2 $R_\textrm{Jup}$ planet has $\log(g) = 2.8$, which is below the bounds of all our model grids. If we instead use the mass posterior for PDS 70 b from Section \ref{sec:dynamics} as a prior on $\log(g)$, we obtained surface gravity values for PDS 70 b near the lower bound of all of the model grids, but none of the other atmospheric parameters changed significantly. As spectroscopic masses from surface gravity and radius have been shown to be unreliable for brown dwarf atmospheres of comparable temperatures \citep[e.g.,][]{Zhang2020}, we avoid overinterpreting the results on these protoplanetary photospheres.

It appears that the 1-5 $\mu$m data alone does not provide significant evidence CPD emission.
Evidence for a second blackbody component by itself is marginal in our fits. For both PDS 70 b and c, the models with a second blackbody component for both the blackbody and DRIFT-PHOENIX models have smaller $B$ than the plain models. Adding a second blackbody does improve the Bayes factor from the plain models for the BT-SETTL and Exo-REM models for both planets, but only the BT-SETTL model for PDS 70 b with a hot compact second component has an adequate Bayes factor. 

The addition of extinction to the BT-SETTL and Exo-REM atmospheric models combined with the second blackbody component generally improved the Bayes factor significantly more than the addition of the second blackbody component alone. The BT-SETTL model with extinction and a second blackbody component has the highest Bayes factor for all of the models considered for PDS 70 b. However, all other extincted models with a second blackbody result in a lower Bayes factor than those with the addition of just ISM extinction alone.

Switching from the pure blackbodies to accreting CPD models from \citet{Zhu2015} only decreases $B$, so there is no evidence that these models are better. The $M_p \dot{M}$ we derived are consistent with mass and mass-accretion values from evolutionary models \citep{Wang2020}. At these low mass accretion rates, the CPD SEDs look similar to blackbody emission \citep{Zhu2015}, but may be less flexible than the blackbody model (e.g., the blackbody model prior range is flexible enough that it can negligibly alter the planetary SED in the observed spectral ranges if needed), resulting in a worse fit. 

Evaluation of CPD emission would not be complete without considering emission at 855 $\mu$m from PDS 70 c, which argues for circumplanetary dust emission from PDS 70 c \citep{Isella2019}. This data point has a significant impact on the evidence for CPD emission given its large spectral lever arm. Unlike in the previous case of considering just 1-5 $\mu$m data, the DRIFT-PHOENIX model augmented with a cooler blackbody has the highest evidence by a factor of $10^4$, strongly ruling out models without a CPD (as seen in Table \ref{table:c_alma_forwardmodels}). Reassuringly, the parameters of the atmospheric model remain unchanged from Table \ref{table:c_forwardmodels}, demonstrating that this 855 $\mu$m data point only probes CPD emission. Thus, our previous conclusions regarding the atmospheric properties of these protoplanets using soley the 1-5 $\mu$m data should hold. With this single data point constraining the CPD properties, the radius and temperature of the second blackbody component are dengenerate, but are consistent with the values found by \citet{Isella2019}.

Thus, we found that there is some evidence for a second blackbody component for both planets when only considering the 1-5 $\mu$m data. The inclusion of the ALMA 855 $\mu$m detection for PDS 70 c definitively rejects models without a second blackbody component, demonstrating the need for observations at longer wavelengths to characterize the circumplanetary dust. These findings are consistent with those of \citet{Stolker2020}, who found that the sole driver of the second blackbody model for PDS 70 b was their $M$-band photometry point, and \citet{Isella2019}, who originally presented to detection of the CPD around PDS 70 c at 855 $\mu$m.

\subsection{What emission are we seeing?}
We interpret the favoring of planetary atmosphere models over featureless blackbody models to indicate that we indeed are seeing into the atmospheres of these protoplanets and that the accreting dust is not completely blocking all molecular signatures as was proposed by \citet{Wang2020}. The effective radii of PDS 70 b from the best-fitting BT-SETTL model with extinction and an added blackbody component is between 1.8 and 2.2 $R_\textrm{Jup}$. Similarly for PDS 70 c, the plain DRIFT-PHOENIX model inferred radii between 1.7 and 2.3 $R_\textrm{Jup}$. Regardless, these effective radii are much smaller than has been found from previous works \citep{Muller2018, Wang2020}, and are even starting to be consistent with hot-start evolutionary models \citep{Baraffe2003}. Using the protoplanetary evolutionary models from \citet{GinzburgChiang2019}, for this age and luminosity, these effective radii are consistent with the lowest mean opacities for the atmospheres of the protoplanets ($< 2 \times 10^{-2}~\textrm{cm}^{2}/\textrm{g}$ using the values for PDS 70 b). Such low opacities have been predicted to occur due to the accretion of dust grains that have undergone grain growth or by coagulation of grains after accretion onto the planet, resulting in a distribution of grain sizes that favors more larger-sized grains than typical ISM distributions \citep{Mordasini2014, Piso2015}. 

The plain DRIFT-PHOENIX model without any modifications has some of the highest Bayes factors for both planets. Given that these models were originally designed to fit dusty, but older, brown dwarfs, we investigated why they have the most support from the data we have obtained so far. We found that this is due to the fact the DRIFT-PHOENIX models do not reproduce the L-T transition by having the clouds clear up at lower effective temperatures, but rather produce thicker clouds at $T_\textrm{eff} < 1600$~K that create a redder near-infrared spectrum \citep{Witte2011}. Given that our retrieved $T_\textrm{eff}$ are all lower than 1600~K, all of the best-fit plain DRIFT-PHOENIX models should appear more dusty than typical substellar atmospheres due to the model creating a large dust cloud purely from atmospheric physics. However, the PDS 70 planets are known to be accreting \citep{Haffert2019}, with the accreting dust expected to shroud the atmosphere \citep{Wang2020}. Thus, we caution against interpreting these model-fitting results as indicating that PDS 70 b and c are just extremely cloudy substellar objects. Instead, it may be that this known deficiency in the DRIFT-PHOENIX models of producing extremely cloudy planets for $T_\textrm{eff} < 1600$~K may be emulating extinction from accreting dust to current measurement precision.

The plain DRIFT-PHOENIX models may not be so different from the extincted BT-SETTL and Exo-REM models that also have significant support from the data. These extincted models also seek to redden the planetary atmosphere to better match the overall 1-5~$\mu$m SED of both planets. The similar extinction amplitudes with dusty brown dwarfs that are not actively accreting may be a coincidence due to large uncertainties in the extinction characteristics. The sedimentation timescale of the dust in these protoplanet atmospheres should be on the order of $10$~years \citep{Wang2020}, so it is unlikely that we are seeing lingering dust from accretion in the field brown dwarfs. The formation of aerosols in the upper atmosphere through some undetermined process has been proposed to explain the dusty brown dwarf population instead \citep{Hiranaka2016}.

Overall, we interpret the fact that DRIFT-PHOENIX models and extincted BT-SETTL and Exo-REM models having the most support form the data to indicate that the planetary atmospheres are indeed significantly extincted by dust from the the planet formation process.
The current spectral data does not well constrain the dust properties, so it is difficult to say how consistent the dust properties are compared to the $10^{-2} ~\textrm{cm}^2/{g}$ that is implied by evolutionary models. However, the $A_V$ required for the extincted BT-SETTL and Exo-REM models are consistent with the constraint that $A_{H\alpha} > 2$~mag to be consistent with the non-detection of either planet with $H\beta$ spectroscopy \citep{Hashimoto2020}. The drastically higher extinction of $A_V\sim10$ for PDS 70 c compared to PDS 70 b implies significantly more dust shrouding PDS 70 c. This could be inherent to the planet or circumplanetary environment since only PDS 70 c has a significant mm signal at its position, which implies a larger CPD than PDS 70 b \citep{Isella2019}. On the other hand, the inferred accretion rates are lower for PDS 70 c, which would imply that it harbors less dust around it \citep{Haffert2019, Wang2020}. Alternatively, the extinction could be enhanced due to additional extinction by the flared edge of the circumstellar disk \citep{Keppler2018}, which has been ignored in this and previous studies of PDS 70 c \citep{Mesa2019, Wang2020}. Better characterization of the 3D vertical structure of the circumstellar disk can help pinpoint if the extinction is due to the circumplanetary or circumstellar environment.

\subsection{Br$\gamma$ Upper Limits}
Although both protoplanets have been seen to emit in H$\alpha$ \citep{Wagner2018,Haffert2019}, no discernible Br$\gamma$ emission has been seen in previous observations \citep{Christiaens2019MNRAS} or in our GRAVITY spectra. Here, we quantified some upper limits on the Br$\gamma$ luminosity and its constraints on the accretion rates. 

We can decompose the planetary flux density that we measure into continuum and line emission:
\begin{equation}
    F_\planet(\lambda) = F_{\planet,\textrm{cont}}(\lambda) + F_{\planet,\textrm{Br}\gamma}(\lambda).
\end{equation}
The continuum emission $F_{\planet,\textrm{cont}}$ is the broad $K$-band spectral shape that we have measured in our GRAVITY spectrum and analyzed in the previous SED fitting sections.
Since the H$\alpha$ emission does not appear to be resolved at 10 times higher spectral resolution than our GRAVITY observations, we expect that any Br$\gamma$ emission would be unresolved with a spectral shape that is simply the line spread function of the GRAVITY instrument centered at the Br$\gamma$ line. Assuming a Gaussian line spread function $LSF(\lambda)$ with standard deviation $\sigma_{LSF}$, the flux density of the planet's line emission $F_{\planet,\textrm{Br}\gamma}$ can be written as
\begin{equation}
\begin{split}
     F_{\planet,\textrm{Br}\gamma}(\lambda) &= \frac{f_{\planet,\textrm{Br}\gamma}}{\sqrt{2\pi}\sigma_{LSF}} \exp\left(-\frac{(\lambda - \lambda_{Br\gamma})^2}{2 \sigma_{LSF}^2} \right) \\
     &\equiv f_{\planet,\textrm{Br}\gamma} LSF_{Br\gamma}.
\end{split}
\end{equation}
Here, $f_{\planet,\textrm{Br}\gamma}$ is the flux of the Br$\gamma$ line integrated over all wavelengths, $\lambda_{Br\gamma} = 2.166 \mu$m, and $\sigma_{LSF} = 0.0018~\mu$m. 

We estimated the integrated flux of the Br$\gamma$ line by performing a matched filter of the continuum subtracted GRAVITY spectra with $LSF_{Br\gamma}$, the expected spectral shape of Br$\gamma$ emission line. The matched filter across the GRAVITY spectral channels is written as:
\begin{equation}
    f_{\planet,\textrm{Br}\gamma} = \frac{ (F_\planet - F_{\planet,\textrm{cont}})^T C^{-1} (LSF_{Br\gamma})}{(LSF_{Br\gamma})^T C^{-1} (LSF_{Br\gamma})},
\end{equation}
where $C$ is the covariance matrix of $F_\planet(\lambda)$ that we computed as part of our spectral extraction in Section \ref{sec:redSpectra}.  The error on $f_{\planet,\textrm{Br}\gamma}$ is then defined as
\begin{equation}
    \sigma_{f,\textrm{Br}\gamma} = \frac{1}{\sqrt{(LSF_{Br\gamma})^T C^{-1} (LSF_{Br\gamma})}}.
\end{equation}

We approximated the continuum emission $F_{\planet,\textrm{cont}}$ by applying a 41-channel median filter on the original planetary spectrum, $F_\planet$. Subtracting off this continuum emission from the original spectrum of each planet gave us an estimated wavelength-integrated Br$\gamma$ flux of $0.1 \pm 1.7 \times 10^{-20}~\textrm{W}/{\textrm{m}^2}$ for PDS 70 b, and  $0.3 \pm 1.3 \times 10^{-20}~\textrm{W}/{\textrm{m}^2}$ for PDS 70 c. Both are fully consistent with non-detections. These correspond to 3$\sigma$ upper limits of $5.1 \times 10^{-20}~\textrm{W}/{\textrm{m}^2}$ for PDS 70 b and $4.0 \times 10^{-20}~\textrm{W}/{\textrm{m}^2}$ for PDS 70 c. The PDS 70 b upper limit is consistent with the 5$\sigma$ upper limit of $8.3 \times 10^{-20}~\textrm{W}/{\textrm{m}^2}$ derived by \citet{Christiaens2019MNRAS}.

We also followed the accretion luminosity and mass accretion analysis from \citet{Christiaens2019MNRAS}, acknowledging that the relations are not calibrated to planetary mass companions forming in the disk, so unknown biases exist. Using the \cite{Calvet2004} relation between Br$\gamma$ and total accretion luminosity for T Tauri stars, we find upper limits on the total accretion luminosity of $<9.6 \times 10^{-5} L_\odot$ and $<7.7 \times 10^{-5} L_\odot$ for PDS 70 b and PDS 70 c, respectively. Then, we can relate total accretion luminosity ($L_\textrm{acc}$) to mass accretion rate $\dot{M}$ using the equation $\dot{M} = 1.25 L_\textrm{acc} R_\planet / G M_\planet$ that \citet{Christiaens2019MNRAS} used from \citet{Gullbring1998}. For PDS 70 b, assuming a mass of 3~$M_\textrm{Jup}$ and a radius of 2~$R_\textrm{Jup}$ based on evolutionary model fits \citep{Wang2020}, we found an upper limit on the mass accretion rate of $< 2.9 \times 10^{-7} M_\textrm{Jup}/$yr for PDS 70 b. For PDS 70 c, assuming a mass of 2~$M_\textrm{Jup}$ and a radius of 2~$R_\textrm{Jup}$ from the same evolutionary models, we found an upper limit on the mass accretion rate of $< 3.4 \times 10^{-7} M_\textrm{Jup}/$yr. Both rates are consistent with most literature results \citep{Wagner2018, Haffert2019, Christiaens2019MNRAS, Wang2020}. This PDS 70 b upper limit from Br$\gamma$ is incompatible with the lower limit derived by \citet{Hashimoto2020} based on the detection of H$\alpha$ but non-detection of H$\beta$, and only marginally consistent with the range of mean accretion rates derived by \citet{Wang2020} from evolutionary models. The disagreements are not surprising given that we used relations that were calibrated to stars and not planets. Models of planetary accretion are needed to translate our Br$\gamma$ upper limits to more realistic mass accretion limits.

\section{Spatially Resolving the Circumplanetary Environment}\label{sec:circum}

\begin{figure*}
    \centering
    \includegraphics[width=0.85\textwidth]{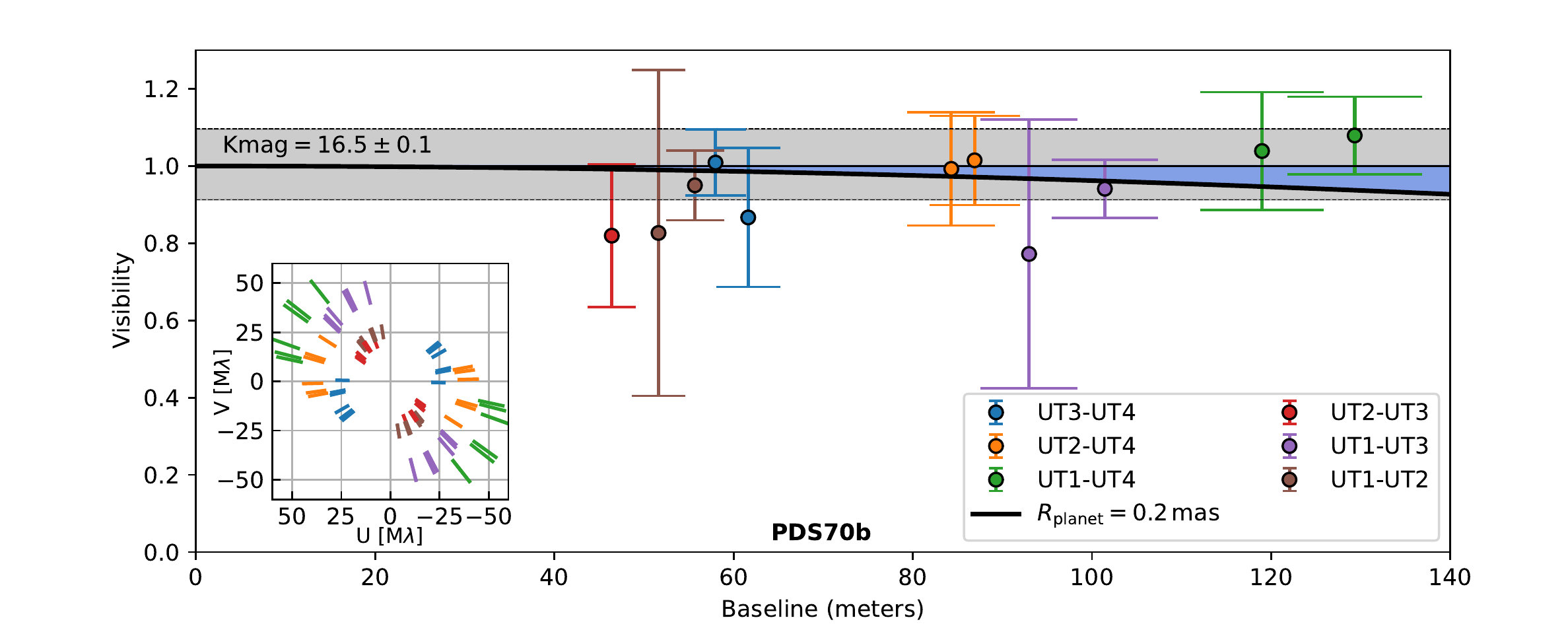}
    \includegraphics[width=0.85\textwidth]{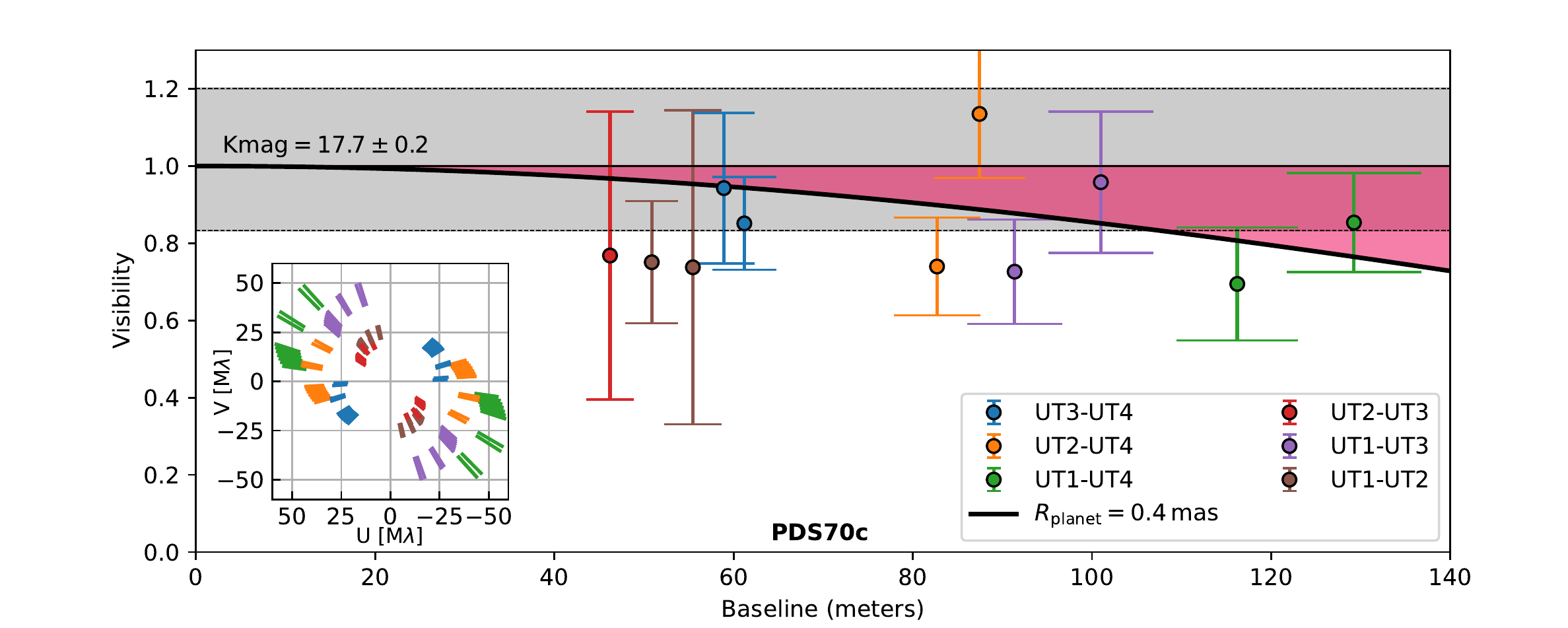}
    \caption{Normalized visibilities ($J_\planet(b)$) of planets b and c as a function of projected baseline length. The data from each night are averaged by baseline. A visibility of 1 corresponds to a coherent flux equal to the flux measured by SPHERE direct imaging. The uncertainty on the normalization due to errors in the SPHERE $K$-band photometry are shown as the gray-shaded region. The blue- and red-shaded regions represent the range of model visibilities allowed based on 1$\sigma$ upper limits on the radii of PDS 70 b and c respectively, assuming all the emission is coming from a uniform disk}. Inset panels: u-v plane coordinates.
    \label{fig:vis}
\end{figure*}

\subsection{VLTI Capabilities}

With baselines up to 130~m, the VLTI at $K$-band wavelengths can achieve an angular resolution ($\lambda/2B$) of $\sim$2\,mas (and can marginally resolve features at smaller angular scales given a sufficient signal-to-noise ratio in the data). The PDS\,70 planetary system has a parallax of 8.8\,mas \citep{Gaia2018}, meaning that GRAVITY is able to spatially resolve scales down to at least 0.2 au (400 $R_\textrm{Jup}$) in projected separation. Here, we present our attempt to resolve the circumplanetary environments of the protoplanets.

Qualitatively, the marker of a resolved circumplanetary environment would be a drop in the coherent flux coming from the planet and its circumplanetary environment. We can look at two indicators to assess whether any emission was spatially resolved. The first indicator is the total coherent flux measured by GRAVITY, which probes spatial scales of a few mas, compared to the total flux within $\sim$40 mas measured by single-dish telescopes in the $K$ band. If GRAVITY is spatially resolving the source, the coherent flux should be lower than the incoherent flux. With uncertainties of $\sim$10-20\% from SPHERE photometry, we did not see a significant drop for either PDS 70 b or c. 

The second indicator would be a drop of the coherent flux as a function of increasing baseline. It would indicate that the longest baselines are spatially resolving structure that appears point like to the shorter baselines. Again, we did not see any significant drop of at least $\sim$10\% in the flux going to longer baselines (see Figure \ref{fig:vis}). However, we used this technique to quantify upper limits on the spatial extent of the circumplanetary region. This is done by fitting synthetic visibility models to the measured visibilities.

\subsection{Uniform Disk Model}

In this section, we considered the case where all the $K$-band emission is coming from a uniform circular disk. This could be true early on in the planet's formation, when the planet has just finished the runaway accretion phase and could have radii of $\sim$1000~$R_\textrm{Jup}$ or $\sim$0.5~au \citep{GinzburgChiang2019}. This is unlikely for these planets, given that the photometrically derived radii (see Section \ref{sec:sed}) are orders of magnitude smaller, and that the mass accretion rates measured with H$\alpha$ and through evolutionary models have indicated that these planets are near the end of their formation process when their radii have contracted to near their final radii \citep{Wagner2018, Haffert2019, Wang2020}. Alternatively, this could approximate the case where the emission is dominated by the CPD \citep{Zhu2015, Szulagyi2019}, although in Section \ref{sec:sed_model_comp} we favored emission from the planetary atmosphere. Regardless, we considered this a simple and limiting case for our ability to resolve the planet or its circumplanetary environment.

In the previous sections, we assumed that each planet was point like. This assumption resulted in Eq.~(\ref{eq:visPlanet}) where the contrast is equal to the planetary flux. We can remove this assumption by including a normalized visibility term $J_\planet(b,t,\lambda)$ proportional to the ratio between coherent flux, $V_\planet(b,t,\lambda)$, and total flux from the protoplanetary object, $F_\planet(\lambda)$:
\begin{equation}
  V_\planet(b, t, \lambda) = J_\planet(b,t,\lambda) F_\planet(\lambda) \,.
  \label{eq:visPlanet2}
\end{equation}
The normalized visibility can therefore be obtained from the ratio of the coherent flux between the planet and star, $R(\lambda,b,t)$, using the equation:
\begin{equation}
  J_\planet(b,t,\lambda)  = R(\lambda,b,t) \frac{ J_\refstar(b, t, \lambda) F_\refstar(\lambda)  }{ \rho \ F_\planet(\lambda) } \,,
\end{equation}
where $J_\refstar(b, t, \lambda)$, $F_\refstar(\lambda)$, and $\rho$ are established in Section~\ref{sec:redSpectra}.
Additionally, it requires the knowledge of $F_\planet(\lambda)$. We assumed in this section that $F_\planet(\lambda)$ is a BT-SETTL model of temperature 1300\,K. The exact model used is not important given our measurement precision. What plays a major role is the absolute scaling of the model: the higher it is, the lower the normalized visibility of the planet will be. In this work, we scaled the model to agree with the $K$-band magnitude of $16.5\pm0.1$ for PDS\,70\,b \citep{Muller2018} and $17.7\pm0.2$ for PDS\,70\,c \citep{Mesa2019} that were obtained through imaging. 

The visibilities as a function of projected baseline length are plotted in Figure~\ref{fig:vis}.
To increase the signal-to-noise ratio, we averaged the visibility $J_\planet(b,t,\lambda)$ over all wavelengths and time:  only one visibility value is obtained for each epoch and baseline. 
We used a uniform disk model, a Bessel function of first order ($\rm J_1$) with $R_\planet$ as the exoplanet radius and $\sqrt{u^2+v^2}$ the projected length of the baseline in units of $\lambda$:
\begin{equation}
J_\planet(b,t,\lambda)=\frac{{\rm J}_1\left(2  \pi R_\planet \sqrt{u^2+v^2}\right) }{  \pi R_\planet \sqrt{u^2+v^2}} \,.
\end{equation}
Fitting these models to both PDS\,70\,b and c, we derived a 1$\sigma$ upper limit on the radii of the exoplanets of, respectively, 0.2 and 0.4\,mas. Using a 3$\sigma$ upper limit, we found a maximum planetary radius of 143~$R_{\rm Jup}$ for PDS 70 b and 285~$R_{\rm Jup}$ for PDS 70 c. These upper limits are consistent with our photometrically derived radii from Section \ref{sec:sed} and protoplanet evolutionary models that are much smaller \citep{GinzburgChiang2019}.

\subsection{Circumplanetary Disk Model}

In reality, we expect an exoplanet of size $\approx1-2\,R_{\rm Jup}$ with a circumplanetary disk (CPD) that is more extended. Thus, we constructed a two-component model where a small uniform disk represents the planet, and a larger uniform disk represents the CPD. For simplicity, we fix the radius of the planet to be 2~$R_\textrm{Jup}$, which is consistent with our photometrically derived radii from Section \ref{sec:sed}.
We also assumed that the CPD contributed 10\% of the $K$-band flux. This is near the upper bound of what is predicted by our models that include a second blackbody component in the SED fits. Alternatively, the CPD could be bright due to scattered starlight, since scattered starlight dominates the $K$-band emission of the outer circumstellar disk. Regardless, a CPD brightness much lower than this would be completely undetectable given our measurement errors, so our analysis here is only suitable for constraining a bright CPD. Thus, we created a model which included the contribution of both the planet and the CPD:
\begin{equation}
J_\planet(b)=0.9 \frac{2{\rm J}_1(x_\planet ) }{ x_\planet  }+ 0.1\frac{2{\rm J}_1(x_{\rm CPD}) }{x_{\rm CPD}}
\end{equation}
where the only free parameter is the radius of the CPD ($R_\textrm{CPD}$):
\begin{eqnarray}
x_\planet &= &2  \pi R_\planet \sqrt{u^2+v^2}\\
x_{\rm CPD} &= & 2  \pi R_{\rm CPD} \sqrt{u'^2+v'^2}
\end{eqnarray}
Note that the projected baseline length, $\sqrt{u'^2+v'^2}$, which corresponds to a position in the frequency plane, is modified to account for the inclination of the CPD. The CPD will look shorter in the direction perpendicular to its position angle due to viewing geometry. We assumed the CPD has the same orientation and inclination as the circumstellar disk. We took values derived from ALMA mm measurements: the inclination of the disk is $i=51.7^\circ$ and the position angle is $\alpha=156.7^\circ$ \citep{Keppler2019}. Hence, the u-v plane is shifted to the new coordinates:
\begin{eqnarray}
u'&=&\cos i \ ({u \cos \alpha-v \sin \alpha})  \\
v'&=&u \sin \alpha + v \cos \alpha \,.
\end{eqnarray}
For PDS\,70\ b, we found a 1$\sigma$ upper limit of $R_{\rm CPD} < 1\,$mas, which we translated to a 3$\sigma$ upper limit of 0.3 au. Although we did not detect any signature of a CPD, we have demonstrated the first observations that can resolve scales of $< 1$~au around a protoplanet.

For PDS\,70\ c, the fit is dominated by the error caused by the normalization of the visibilities (the visibility of 1 is assumed to correspond to a coherent flux of ${\rm Kmag} = 17.7\pm0.2$). If we neglected this uncertainty, the fit converged to a large CPD of radius $R_{\rm CPD} > 10\,$mas. Because the extent of such CPD seems unrealistic (see below), we are inclined to think that the drop in visibility is purely instrumental, due to the uncertainty in the SPHERE $K$-band magnitude.

\subsection{Comparison to Predicted Circumplanetary Disk Sizes}

The size of a CPD is governed by either the Hill or Bondi radius, whichever is smaller. The Hill radius ($R_H$) for an eccentric planet is defined as
\begin{equation}
    R_H = a ( 1-e) \sqrt[3]{M_p/(3 M_*)},
\end{equation}
where $a$ is the semi-major axis, $e$ is the eccentricity, $M_p$ is the planet mass, and $M_*$ is the stellar mass. For PDS 70 b, if we use a planet mass of 3 $M_\textrm{Jup}$ based on evolutionary models \citep{Wang2020}, a stellar mass of 0.9 $M_\odot$, a semi-major axis of 20~au, and an eccentricity of 0.2 from our orbit fit (Section \ref{sec:orbit}), then we derive $R_H = 1.6$~au. 

The Bondi radius for an isothermal ideal gas is
\begin{equation}
    R_B= \frac{G M_p}{c_s^2} = \frac{G M_p \mu}{k_B T},
\end{equation}
where $\mu$ is the mean molecular weight, $G$ is the gravitational constant, $k_B$ is the Boltzmann constant, $M_p$ is the planet mass, $T$ is the gas temperature, and $c_s$ is the sound speed. Assuming the gas is a blackbody in thermal equilibrium with the star, the gas temperature is 50~K. Assuming the same planet mass as before and $\mu = 2$~amu for molecular hydrogen, we found a Bondi radius of 8~au. Our upper limit of $R_\textrm{CPD} < 0.3$\,au thus corresponds to 0.2~$R_H$ or 0.04~$R_B$. 

With $R_H < R_B$, we are likely looking at a planet that is slightly above the thermal mass, the mass above which accretion starts transitioning into a 2D process from a 3D one \citep{Ginzburg2019a, Rosenthal2020}.
We compared our upper limit on PDS 70 b's CPD to the CPD models of \citet{Szulagyi2017} and \citet{Fung2019} that explore planet masses above the thermal mass regime. \citet{Szulagyi2017} reported results in $R_H$ whereas \citet{Fung2019} provided disk sizes in $R_B$. For planets with comparable Hill radii, \citet{Szulagyi2017} found CPDs with spatial scales of up to 0.1~$R_H$, which agrees well with our upper limit of 0.2~$R_H$. \citet{Fung2019} found that the CPD are about $0.1~R_B$ for planets just above the thermal mass. We found a upper limit in units of Bondi radii that is $\sim3$ times smaller. We note though that we assumed a very bright CPD (10\% of the total $K$-band flux), and that a fainter CPD would escape detection even if it was more extended than $0.04~R_B$. Our upper limit of 0.3~au is also consistent with the upper limit of 0.1~au for a CPD around PDS 70 b derived using SED fitting by \citet{Stolker2020}, accounting for a non-detection at 855 $\mu$m.

For PDS 70 c, using a planet mass of 2 $M_\textrm{Jup}$ \citep{Wang2020}, a semi-major axis of 30~au, and an eccentricity of 0, we found $R_H = 2.7$~au. Assuming a gas temperature of 40~K at 30~au, we find $R_B = 10$~au. The limits from \citet{Szulagyi2017} and \citet{Fung2019} imply CPD sizes of 0.5~au (5~mas) and 1~au (9 mas) respectively. Both are smaller than the 10~mas mentioned in the previous suction that would explain the visibility normalization offset, arguing for photometric calibration to be responsible for the offset.

\section{Conclusions}\label{sec:conclusions}
In this work, we present interferometric observations of protoplanets PDS 70 b and c, as well as their host star, using the GRAVITY instrument at VLTI. Using baselines up to 130~m at $K$ band, we obtained the highest spatial-resolution observations of the system to date. We spatially resolved the inner circumstellar disk, finding that it contributes at least 6\% of the $K$-band luminosity from the system. Such an excess is consistent with SED fits of the star that only use photometry with wavelengths shorter than the $K$ band that find a 14\% excess. It is uncertain whether the remaining 8\% emission is further out than 6 au, or closer in than 0.2 au.

We obtained R$\sim$500 $K$-band spectra and 100 $\mu$as astrometry of both protoplanets over two epochs. 
We fit the GRAVITY astrometry on both planets to a near-coplanar orbital model and rejected orbits that are not dynamically stable for 8~Myr. We found a nonzero eccentricity for PDS 70 b of $0.17 \pm 0.06$, whereas the orbit of PDS 70 c is nearly circular. The semi-major axes and eccentricities are consistent with models of the two planets migrating into 2:1 MMR while accreting from the circumstellar disk \citep{Bae2019}, but we found that MMR may not be required for dynamical stability. Agnostic to whether the planets have to be in MMR, we placed a dynamical mass upper limit of 10~$M_\textrm{Jup}$ for PDS 70 b that is consistent with predictions from evolutionary models \citep{Wang2020}. We were not able to constrain the mass of PDS 70 c dynamically. However, we were able to constrain the mass of the star to be $0.982 \pm {0.066}~M_\odot$, which is 1.6$\sigma$ higher than the masses derived from the stellar SED fits in this paper and dynamical mass measurements from velocity maps of the circumstellar gas \citep{Keppler2019}. Future GRAVITY astrometry will constrain the orbital acceleration of the planets at GRAVITY-level precision and will significantly improve our orbital and mass constraints.

We combined our GRAVITY $K$-band spectra of both places with a re-reduction of the SPHERE IFS spectrum of PDS 70 c and archival data to characterize the photospheres of both planets.
We considered four atmospheric forward models, each with a suite of modifications to account for extinction and circumplanetary dust emission, to describe the photospheres of the two protoplanets. The spectral shape of the GRAVITY $K$-band spectra were able to reject pure blackbody models for both planet unlike previous work \citep{Wang2020,Stolker2020}. We found the best fitting models are plain DRIFT-PHOENIX models or extincted BT-SETTL and Exo-REM models. Both classes of models appear to be emulating a dusty planetary atmosphere that can arise if accreting dust shrouds the atmospheres of the protoplanets. However, we cannot pinpoint the location of the dust (e.g., in the atmosphere, around the planet, in the circumplanetary or circumstellar disk) with our analysis. The extinction values we found for the dust are consistent with the non-detection of H$\beta$ \citep{Hashimoto2020}. 
The fact we favored planetary atmosphere models is promising as better observations of these protoplanet atmospheres may be able to constrain their atmospheric composition, which can then be related to measurements of the composition of circumstellar material. PDS 70 is currently the only system that can allow us to directly study how the final composition of a planet comes to be. 

When compared to evolutionary modes, the inferred photospheric radii of $1.7-2.3$~$R_\textrm{Jup}$ implied that the dust grains have low mean opacities of $< 2 \times 10^{-2}~\textrm{cm}^2/\textrm{g}$, and could be evidence for grain growth \citep{GinzburgChiang2019}. We also placed upper limits on Br$\gamma$ emission of $< 5.1 \times 10^{-20}~\textrm{W}/{m^2}$ for PDS 70 b and $< 4.0 \times 10^{-20}~\textrm{W}/{m^2}$ PDS 70 c from our GRAVITY spectra. 

With 1-5 $\mu$m spectrophotometric data alone, the evidence for CPDs is not definitive. We found some evidence for seeing emission from a circumplanetary disk in PDS 70 b, but more data at longer wavelengths are necessary to confirm such a hypothesis, as the current findings rely on the single $M$-band photometry point from \citet{Stolker2020}. Only when including the 855 $\mu$m detection of continuum emission from PDS 70 c from \citep{Isella2019} did we find definitive evidence to reject models without a CPD for PDS 70 c. However, this data point alone cannot constrain both the temperature and radius of the CPD, as none of the our 1-5 $\mu$m data provided significant constraints.  

With an angular resolution of 2~mas (0.2 au), we were able to spatially probe the circumplanetary environment of the protoplanets. We did not find any evidence that we spatially resolved either protoplanet or its CPD. Assuming that all the emission is coming from a uniform sphere, we placed 3$\sigma$ upper limits on the radius of the sphere to be 285 and 499 $R_\textrm{Jup}$ for PDS 70 b and c, respectively. Alternatively, we considered a model with a compact photosphere with a radius of 2 $R_\textrm{Jup}$ emitting 90\% of the $K$-band flux and a bright, extended disk emitting the rest. We placed an upper limit on the size of the bright CPD of 0.1 au for PDS 70 b that corresponds to 0.2 $R_H$ or 0.04 $R_B$, consistent with models of CPDs \citep{Szulagyi2017, Fung2019}. We note that fainter diffuse emission further out would have escaped detection. Larger infrared interferometer arrays, like the proposed Planet Formation Imager, are needed to spatially resolve the CPDs around these planets \citep{Monnier2018}. 

\acknowledgments

We thank Dino Mesa and Michael Liu for helpful discussions.
This research has made use of the Jean-Marie Mariotti Center \texttt{Aspro}
service.
J.J.W., S.G., P.G., and S.B. acknowledge support from the Heising-Simons Foundation, including grant 2019-1698.
AV acknowledges funding from the European Research Council (ERC) under the European Union’s Horizon 2020 research and innovation programme (grant agreement No. 757561).
T.H., P.M and R.A-T acknowledge support from the European Research Council under the Horizon 2020 Framework Program via the ERC Advanced Grant Origins 83 24 28.
SPHERE is an instrument designed and built by a consortium consisting of IPAG (Grenoble, France), MPIA (Heidelberg, Germany), LAM (Marseille, France), LESIA (Paris, France), Laboratoire Lagrange (Nice, France), INAF - Osservatorio di Padova (Italy), Observatoire de Gen\`eve (Switzerland), ETH Z\"urich (Switzerland), NOVA (Netherlands), ONERA (France) and ASTRON (Netherlands) in collaboration with ESO. SPHERE was funded by ESO, with additional contributions from CNRS (France), MPIA (Germany), INAF (Italy), FINES (Switzerland) and NOVA (Netherlands). 
SPHERE also received funding from the European Commission Sixth and Seventh Framework Programmes as part of the Optical Infrared Coordination Network for Astronomy (OPTICON) under grant number RII3-Ct-2004-001566 for FP6 (2004-2008), grant number 226604 for FP7 (2009-2012) and grant number 312430 for FP7 (2013-2016). 
R.G.L. has received funding from Science Foundation Ireland under Grant No. 18/SIRG/5597.

\facility{VLTI (GRAVITY)}

\software{{\tt orbitize!} \citep{Blunt2020}, \texttt{pyKLIP} \citep{Wang2015}, \texttt{vlt-sphere} \citep{Vigan2020ascl}, {\tt emcee} \citep{ForemanMackey2013}, {\tt ptemcee} \citep{Vousden2016}, \texttt{pymultinest} \citep{Buchner2014}, \texttt{REBOUND} \citep{Rein2012,Rein2015}, \texttt{astropy} \citep{Astropy2018} }

\clearpage
\newpage

\bibliography{pds70_clean,pds70_sylvestre,pds70_tomas}

\clearpage

\begin{appendix}

\section{Injection Losses}
\label{sec:losses}

\begin{figure}
    \centering
    \includegraphics[width=0.5\textwidth]{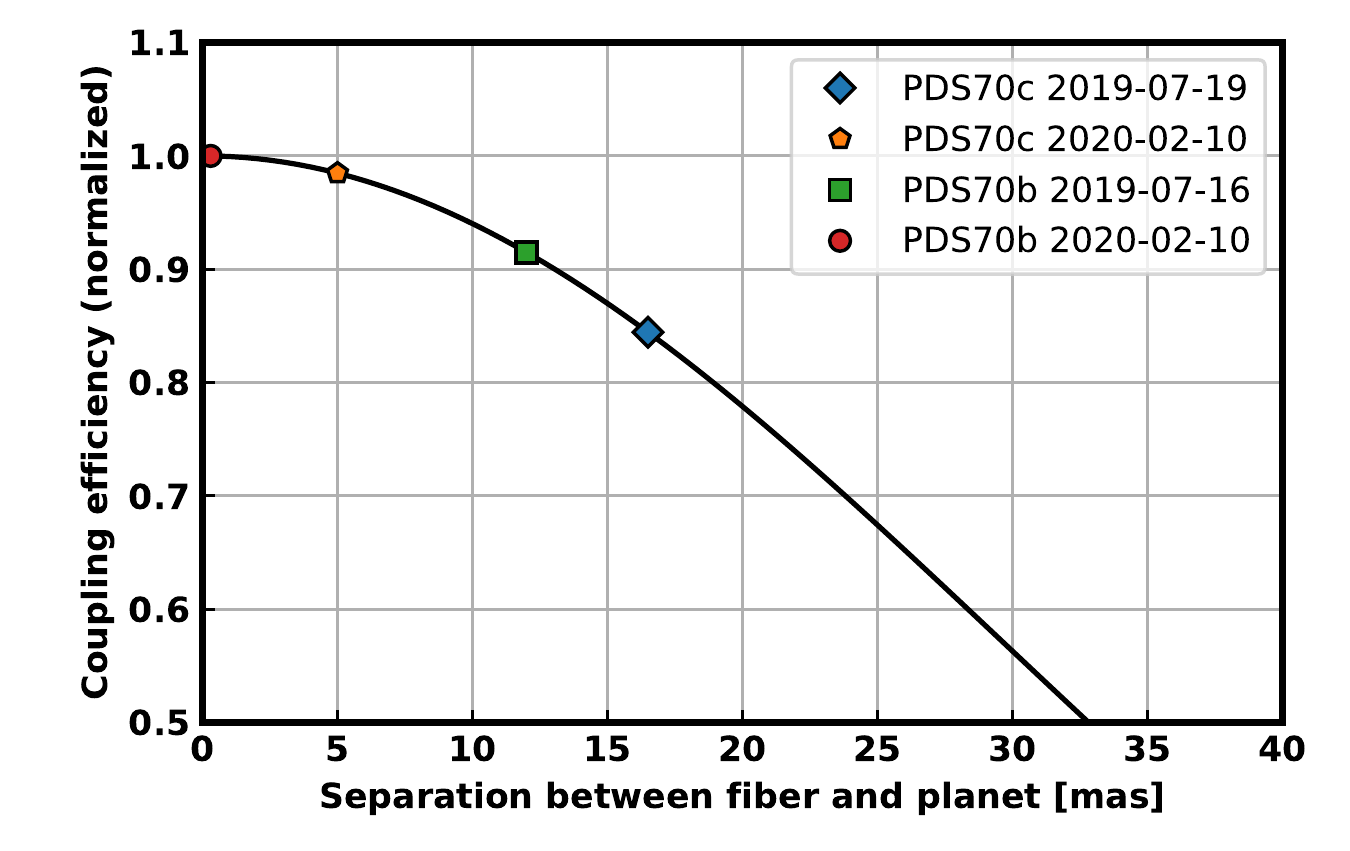}
    \caption{Injection losses $\gamma$ as a function of the misalignment ($\Delta\alpha$) of the fiber. \label{fig:losses}}
\end{figure}

For the first observations of PDS\,70\,b\&\,c, the uncertainty of the position of the exoplanets were high. We positioned the fiber using best-guess positions from previous orbit fits. Therefore, the science fiber was not perfectly centered on the exoplanet. This resulted in injection losses, which we calculated here under the assumption of no atmosphere.

The complex field injected into the fiber is the normalized scalar product between the
 fundamental mode of the fibre ($E$) and the incoming wave front ($U$). According to the Parceval-Planchet theorem, the calculation can be done either
  at the entrance of the fiber (focal plane) or in the Fourier domain (pupil plane). The calculation is easier in the pupil plane. Therefore, using the $x$ and $y$ the coordinates inside the pupil, we can write the complex coupling parameter as:
\begin{equation}
A =
\frac{ \iint_{-\infty}^{+\infty}
  U(x,y)\,.\,E^\star(x,y)\ dx
  dy}{
  \sqrt{
  \iint_{-\infty}^{+\infty}
  |U(x,y)|^2 dx
  dy\,\times\,\iint_{-\infty}^{+\infty}
  |E(x,y)|^2dx dy}} \,.
\end{equation}
The amplitude of the injected flux is $|A|^2$.
In our case, we want to calibrate the losses caused by the fact that the exoplanet flux is slightly coming off axis, with an angle $\Delta\alpha$. Assuming such a tilt, for a telescope of diameter $D$, the electrical field of stellar origin is:
\begin{equation}
U(x,y) \propto
\begin{cases}
\exp{\left(  2 i \pi \Delta\alpha  \frac{x}{ \lambda} \right)} &  \text{ if $\sqrt{x^2+y^2} <D/2$ } \\
0 & \text{otherwise}
\end{cases}
\end{equation}
where the effect of the atmosphere is considered fully corrected by the adaptive optics. The same approach can also include a phase term for the residuals of atmospheric correction \citep{2019A&A...625A..48P}, but the effect of such a term is neglected here.
The fundamental mode of the fiber is a Gaussian profile:
\begin{equation}
E(x,y) \propto {\exp{\left( -  \frac{ x^2 + y^2 }{ 2 \omega_0^2} \right) }}.
\end{equation}
To maximize the injection, the mode is centred on the pupil, and has a half-width maximum of $\omega_0=0.32D$. In such a case, for a full pupil, $|A|^2_{\rm on-axis} = 81\%$.
However, since both observations of the star and the planet were obtained using the same single mode fiber, it is easier to use $\gamma$, the normalized amplitude of the injected flux:
\begin{equation}
  \gamma = \frac{ |A|^2_{\rm misaligned} }{|A|^2_{\rm on-axis}} =
\left|\frac{ \iint_{\rm Pup}
\exp{\left( 2 i \pi \Delta\alpha  \frac{x}{ \lambda} -  \frac{ x^2 + y^2}{ 2 \omega_0^2}  \right) }
dx dy
}{ \iint_{\rm Pup}
\exp{\left( -  \frac{ x^2 + y^2}{ 2 \omega_0^2}  \right) }
dx dy
} \right|^2.
\end{equation}

The result of this calculation is plotted in Figure~\ref{fig:losses} as a function of $\Delta\alpha$. The worst coupling happened for PDS\,70\,c on 2019-07-19, when the fiber was situated 16.5\,mas from the exoplanet. The coupling efficiency was only 84\% of the optimal value (not including the effect of the atmosphere). For the other epoch of c, the exoplanet position was better determined (5\,mas), and the efficiency was 98\%. For the two epochs of PDS\,70\,b, the coupling was 91 and 99\% of the maximum on-axis coupling.

\section{Literature Data}\label{sec:lit_data}

We listed the photometry from the literature for PDS 70 b and PDS 70 c in Table \ref{table:lit_phot}. For each photometric data point, we listed the wavelength ($\lambda$) and half width ($\Delta \lambda$) to represent the wavelength coverage of that filter. We did not list the SPHERE IFS spectra for both planets used in the SED fit.

The literature astrometry for PDS 70 b and PDS 70 c are listed in Table \ref{table:lit_astrom}. These were used to supplement the two GRAVITY epochs for each planet.

\begin{deluxetable*}{c|c|c|c|c|c}
\tablecaption{Literature Photometry \label{table:lit_phot}}
\tablehead{
Planet & \shortstack[c]{ $\lambda$ \\ ($\mu$m)}  & \shortstack[c]{ $\Delta \lambda$ \\ ($\mu$m)} & \shortstack[c]{Flux\\ ($\textrm{W}/\textrm{m}^2/\mu$m)} & \shortstack[c]{Flux Error\\ ($\textrm{W}/\textrm{m}^2/\mu$m)} & Reference
}
\startdata
b & 1.5888 & 0.0266 & $6.43 \times 10^{-17}$ & $1.04\times 10^{-17}$ & \citet{Muller2018} \\
b & 1.5888 & 0.0266 & $7.92 \times 10^{-17}$ & $1.75 \times 10^{-17}$ & \citet{Muller2018} \\
b & 1.6671 & 0.0278 & $6.83 \times 10^{-17}$ & $1.05 \times 10^{-17}$ & \citet{Muller2018} \\
b & 1.6671 & 0.0278 & $7.53 \times 10^{-17}$ & $1.14 \times 10^{-17}$ & \citet{Muller2018} \\
b & 2.100 & 0.051 & $1.203 \times 10^{-16}$ & $3.46 \times 10^{-17}$ & \citet{Muller2018} \\
b & 2.100 & 0.051 & $9.16 \times 10^{-17}$ & $0.43 \times 10^{-17}$ & \citet{Muller2018} \\
b & 2.2510 & 0.0545 & $1.156 \times 10^{-17}$ & $2.57 \times 10^{-17}$ & \citet{Muller2018} \\
b & 2.2510 & 0.0545 & $9.35 \times 10^{-17}$ & $0.42 \times 10^{-17}$ & \citet{Muller2018} \\
b & 3.78 & 0.35 & $7.67 \times 10^{-17}$ & $3.09 \times 10^{-17}$ & \citet{Muller2018} \\
b & 3.80 & 0.31 & $5.95 \times 10^{-17}$ & $2.50 \times 10^{-17}$ & \citet{Muller2018} \\
b & 3.776 & 0.350 & $7.52 \times 10^{-17}$ & $1.22 \times 10^{-17}$ & \citet{Wang2020} \\
b & 4.0555 & 0.0308 & $5.35 \times 10^{-17}$ & $1.36 \times 10^{-17}$ & \citet{Stolker2020} \\
b & 4.7555 & 0.2961 & $6.56 \times 10^{-17}$ & $1.63 \times 10^{-17}$ & \citet{Stolker2020} \\
\hline
c & 2.110 & 0.051 & $3.10 \times 10^{-17}$ & $0.52 \times 10^{-17}$ & \citet{Mesa2019} \\
c & 2.110 & 0.051 & $3.73 \times 10^{-17}$ & $0.40 \times 10^{-17}$ & \citet{Mesa2019} \\
c & 2.251 & 0.055 & $2.77 \times 10^{-17}$ & $0.53 \times 10^{-17}$ & \citet{Mesa2019} \\
c & 2.251 & 0.055 & $3.59 \times 10^{-17}$ & $0.49 \times 10^{-17}$ & \citet{Mesa2019} \\
c & 3.776 & 0.350 & $3.31 \times 10^{-17}$ & $1.31 \times 10^{-17}$ & \citet{Wang2020}
\enddata
\end{deluxetable*}

\begin{deluxetable*}{c|c|c|c|c|c|c}
\tablecaption{Literature Astrometry \label{table:lit_astrom}}
\tablehead{
Planet & \shortstack[c]{ Epoch \\ (MJD)}  & \shortstack[c]{ Separation \\ (mas)} & \shortstack[c]{Sep. Error \\ (mas)} & \shortstack[c]{PA \\ (\degr)} & \shortstack[c]{PA Error \\ (\degr) } & Reference
}
\startdata
b & 56017 & 191.9 & 21.4 & 162.2 & 3.7 & \citet{Muller2018} \\
b & 57145 & 192.3 & 4.2 & 154.5 & 1.2  & \citet{Muller2018} \\
b & 57145 & 197.2 & 4.0 & 154.9 & 1.1  & \citet{Muller2018} \\
b & 57173 & 199.5 & 6.9 & 153.4 & 1.8  & \citet{Muller2018} \\
b & 57173 & 194.5 & 6.3 & 153.5 & 1.8  & \citet{Muller2018} \\
b & 57522 & 193.2 & 8.3 & 152.2 & 2.3  & \citet{Muller2018} \\
b & 57522 & 199.2 & 7.1 & 151.5 & 1.6  & \citet{Muller2018} \\
b & 57540 & 189.6 & 26.3 & 150.6 & 7.1 & \citet{Muller2018} \\
b & 58173 & 192.1 & 7.9 & 147.0 & 2.4  & \citet{Muller2018} \\
b & 58173 & 192.2 & 8.0 & 146.8 & 2.4  & \citet{Muller2018} \\
b & 58289 & 176.8 & 25.0 & 146.8 & 8.5 & \citet{Haffert2019} \\
b & 58642 & 175.8 & 6.9 & 140.9 & 2.2  & \citet{Wang2020} \\
\hline
c & 56017 & 191.9 & 21.4 & 162.2 & 3.7 & \citet{Muller2018} \\
c & 58173 & 209.0 & 13.0 &  281.2 & 0.5 & \citet{Mesa2019} \\
c & 58289 & 235.5 & 25.0 & 277.0 & 6.5 & \citet{Haffert2019} \\
c & 58548 & 225.0 & 8.0 & 279.9 & 0.5 & \citet{Mesa2019} \\
c & 58642 & 223.4 & 8.0 & 280.4 & 2.0 & \citet{Wang2020}
\enddata
\end{deluxetable*}

\end{appendix}

\end{CJK*}
\end{document}